\documentclass[11pt,a4paper]{article}
\ifdefined\pdfoutput\pdfoutput=1\fi
\makeatletter
\IfFileExists{jheppub.sty}{%
  \usepackage{jheppub}%
}{%
  \usepackage[a4paper,width=15.1cm,height=22.6cm,top=3.2cm,left=2.95cm]{geometry}%
  \usepackage{amsmath,amssymb,graphicx}%
  \usepackage{xcolor}%
  \usepackage[colorlinks=true,urlcolor=blue!55!black,anchorcolor=blue!55!black,citecolor=blue!55!black,filecolor=blue!55!black,linkcolor=blue!55!black,menucolor=blue!55!black,linktocpage=true]{hyperref}%
  \numberwithin{equation}{section}%
  \gdef\@jhep@author{}%
  \gdef\@jhep@affil{}%
  \gdef\@jhep@email{}%
  \gdef\@jhep@abstract{}%
  \gdef\@jhep@keywords{}%
  \gdef\@jhep@arxiv{}%
  \renewcommand{\author}[2][]{\g@addto@macro\@jhep@author{##2\if\relax\detokenize{##1}\relax\else\textsuperscript{##1}\fi\par}}%
  \newcommand{\affiliation}[2][]{\g@addto@macro\@jhep@affil{\if\relax\detokenize{##1}\relax\else\textsuperscript{##1}\fi ##2\par}}%
  \newcommand{\emailAdd}[1]{\g@addto@macro\@jhep@email{\textit{E-mail:} \href{mailto:##1}{\texttt{##1}}\par}}%
  \renewcommand{\abstract}[1]{\gdef\@jhep@abstract{##1}}%
  \newcommand{\keywords}[1]{\gdef\@jhep@keywords{##1}}%
  \newcommand{\arxivnumber}[1]{\gdef\@jhep@arxiv{##1}}%
  \newcommand{\acknowledgments}{\section*{Acknowledgments}}%
  \renewcommand{\maketitle}{%
    \thispagestyle{empty}%
    \noindent\rule{\textwidth}{1.2pt}\par
    \vspace{1.2em}%
    {\raggedright\LARGE\bfseries \@title\par}%
    \vspace{1.2em}%
    {\raggedright\large\bfseries \@jhep@author\par}%
    \vspace{0.5em}%
    {\raggedright\normalsize\itshape \@jhep@affil\par}%
    \vspace{0.5em}%
    {\raggedright\small \@jhep@email\par}%
    \vspace{1.5em}%
    \noindent\textsc{Abstract:} \@jhep@abstract\par
    \vspace{1.0em}%
    \ifx\@jhep@keywords\@empty\else
      \noindent\textsc{Keywords:} \@jhep@keywords\par
      \vspace{0.6em}%
    \fi
    \ifx\@jhep@arxiv\@empty\else
      \noindent\textsc{ArXiv ePrint:} \href{https://arxiv.org/abs/\@jhep@arxiv}{\@jhep@arxiv}\par
      \vspace{0.6em}%
    \fi
    \vspace{0.8em}%
    \noindent\rule{\textwidth}{0.6pt}\par
    \tableofcontents
    \vspace{1.2em}%
    \noindent\rule{\textwidth}{0.6pt}\par
    \newpage
  }%
}
\makeatother
\usepackage{orcidlink}

\usepackage{amsthm,mathtools}
\usepackage{tikz}
\usepackage{microtype}

\theoremstyle{plain}
\newtheorem{theorem}{Theorem}
\newtheorem{lemma}{Lemma}
\newtheorem{corollary}{Corollary}
\theoremstyle{definition}
\newtheorem{definition}{Definition}
\newtheorem{assumption}{Assumption}
\theoremstyle{remark}
\newtheorem*{remark}{Remark}

\begin{document}
\title{Python's Lunch hidden in the entanglement shadow}

\author{Libo Jiang\orcidlink{0009-0008-7942-1874} and}
\emailAdd{sustech@buaa.edu.cn}
\author{Yan Liu}
\emailAdd{yanliu@buaa.edu.cn}

\affiliation{Department of Space Science, \\
Beihang University, Beijing 100191, China}
\affiliation{Peng Huanwu Collaborative Center for Research and Education, \\Beihang University, Beijing 100191, China}

\abstract{A Python's Lunch geometry makes decoding information hidden behind a nonminimal extremal surface exponentially hard. Here, we relate this decoding complexity to the entanglement shadow --- the bulk region untouched by any Ryu--Takayanagi (RT) surface. We show that if the complexity of decoding a bulk point exceeds a critical bound, the point must lie inside the entanglement shadow unless the geometry realizes a specific ``throats-on-both-sides'' configuration. We prove this theorem in both the classical RT regime and the semiclassical regime with quantum corrections. Together, these bounds link bulk degrees of freedom that are hard to decode to those hidden from boundary spatial entanglement probes.}


\maketitle

\section{Introduction}
\label{sec:intro}

In AdS/CFT, holographic entanglement entropy --- the Ryu--Takayanagi (RT) prescription~\cite{RyuTakayanagi2006} --- captures key boundary information through a simple geometric picture. Later developments extended this picture to the covariant Hubeny--Rangamani--Takayanagi (HRT) prescription~\cite{HubenyRangamaniTakayanagi2007,Wall2012} and to generalized entropy with bulk matter quantum corrections~\cite{FaulknerLewkowyczMaldacena2013,EngelhardtWall2014}. More recently, entanglement islands reproduced the Page curve of evaporating black holes, shedding light on the black hole information paradox~\cite{Penington2019,AlmheiriEngelhardtMarolfMaxfield2019}. Holographic entanglement entropy thus remains a central tool in holography.

Entanglement alone, however, cannot describe all bulk geometric properties~\cite{Susskind2014}. Computational complexity has therefore been proposed to fill this gap by capturing bulk information invisible to boundary entanglement entropy. For an operation or quantum state, complexity measures the minimal number of basic gates required to implement the operation or prepare the state from a simple reference state. Because complexity continues to grow long after entanglement entropy saturates, it may explain the prolonged growth of the Einstein--Rosen (ER) bridge. Several proposals identify boundary complexity with the volume of a maximal Cauchy slice~\cite{StanfordSusskind2014}, the gravitational action of the Wheeler--DeWitt (WDW) patch~\cite{BrownEtAl2015}, or more general functionals of codimension-one~\cite{BelinEtAl2021} or codimension-zero bulk regions~\cite{BelinEtAl2022}. These holographic complexity measures typically scale at most polynomially in the bulk geometric length scales (or in $1/G_N$, where \(G_N\) is Newton's constant).

This polynomial scaling clashes with the exponential complexity expected for decoding the black hole interior from late-time Hawking radiation~\cite{HarlowHayden2013}. This tension motivates \emph{Python's Lunch} complexity~\cite{BrownGharibyanPeningtonSusskind2019}, which governs a different operational task. While volume complexity counts the total number of tensors in a bulk-to-boundary tensor network, individual tensor layers need not be unitary. Consider a bulk geometry with multiple extremal surfaces, where a locally maximal ``bulge'' lies between two locally minimal ``throats.'' Propagating outward across such a region appears to compress a large Hilbert space into a smaller one, seemingly violating unitarity. Yet the global map from the bulk code subspace to the boundary remains an approximate isometry. Effectively, initial code qubits scramble with auxiliary qubits carrying no independent physical information, jointly spanning a large intermediate Hilbert space at the bulge. Mapping the code space isometrically into a smaller output Hilbert space requires unscrambling the auxiliary qubits and projecting them onto a fixed state. With post-selection, this takes only polynomial circuit complexity proportional to bulk volume; simulating this post-selection with purely unitary operations becomes exponentially hard. Grover's search algorithm (amplitude amplification)~\cite{Grover1996,BrassardEtAl2000} is the fastest known unitary protocol for this task, scaling as
\[
\mathcal C \approx \mathcal C_0\sqrt{\frac{d_{\rm bulge}}{d_{\rm throat}}}
= \mathcal C_0\exp\!\left[\frac{\operatorname{Area}(\text{bulge})-\operatorname{Area}(\text{throat})}{8G_N}\right],
\]
where $d_{\rm bulge}$ and $d_{\rm throat}$ denote the effective Hilbert space dimensions of the bulge and throat, and $\mathcal C_0$ reflects the polynomial complexity of the scrambling circuit. We call this unitary gate count the \emph{decoding complexity}, where nonminimal extremal surfaces are conjectured to be the sole source of exponential complexity in the holographic dictionary~\cite{FindingPythons2021}

This geometric obstruction recalls the \emph{entanglement shadow}~\cite{BalasubramanianEtAl2014,FreivogelEtAl2015} --- the bulk region untouched by the globally minimizing RT surface (or quantum extremal surface, QES) of any boundary subregion. Entanglement shadows arise around $\mathrm{AdS}_3$ conical defects, dense stars, and single-sided black holes, and explicit examples linking bulges to shadows were studied in~\cite{AroraEtAl2024}.

\begin{figure}[t]
\centering
\begingroup
\definecolor{introoutline}{HTML}{617482}
\definecolor{introthroat}{HTML}{354D65}
\definecolor{introbulge}{HTML}{C18A30}
\definecolor{intrort}{HTML}{8B4E83}
\definecolor{introshadow}{HTML}{148779}
\begin{tikzpicture}[x=1.23cm,y=1.15cm,line cap=round,line join=round,
  every node/.style={font=\small}]
\path[use as bounding box] (-.58,-1.98) rectangle (11.00,1.90);
\path[fill=introoutline!7,draw=introoutline,line width=.9pt]
 (0,1.40) .. controls (.28,.72) and (.96,.48) .. (1.90,.48)
 .. controls (2.04,.48) and (2.10,.54) .. (2.26,.54)
 .. controls (2.61,.54) and (2.84,.30) .. (2.84,0)
 .. controls (2.84,-.30) and (2.61,-.54) .. (2.26,-.54)
 .. controls (2.10,-.54) and (2.04,-.48) .. (1.90,-.48)
 .. controls (.96,-.48) and (.28,-.72) .. (0,-1.40) -- cycle;
\draw[introoutline,fill=white,line width=.9pt] (0,0) ellipse (.23 and 1.40);
\draw[introthroat,densely dashed,line width=1pt]
 (1.90,.48) arc[start angle=90,end angle=270,x radius=.10,y radius=.48];
\draw[introthroat,line width=1.3pt]
 (1.90,-.48) arc[start angle=270,end angle=450,x radius=.10,y radius=.48];
\draw[introbulge,densely dashed,line width=1pt]
 (2.26,.54) arc[start angle=90,end angle=270,x radius=.11,y radius=.54];
\draw[introbulge,line width=1.3pt]
 (2.26,-.54) arc[start angle=270,end angle=450,x radius=.11,y radius=.54];
\draw[intrort,densely dashed,line width=1.25pt]
 (-0.199186,-0.700000)
 .. controls (0.073822,-0.670378) and (0.330755,-0.572459) .. (0.600000,-0.500000)
 .. controls (0.964284,-0.401964) and (1.351105,-0.350531) .. (1.700000,-0.320000)
 .. controls (1.968700,-0.296487) and (2.214905,-0.285371) .. (2.550000,-0.200000)
 .. controls (2.687515,-0.164966) and (2.840000,-0.117426) .. (2.840000,0.000000);
\draw[intrort,line width=1.5pt]
 (0.199186,0.700000)
 .. controls (0.458458,0.50132) and (0.696959,0.486078) .. (0.950000,0.420000)
 .. controls (1.262719,0.361262) and (1.597645,0.328016) .. (1.900000,0.300000)
 .. controls (2.105650,0.280944) and (2.296232,0.264308) .. (2.550000,0.200000)
 .. controls (2.685935,0.165552) and (2.840000,0.117426) .. (2.840000,0.000000);
\fill[intrort] (-0.199186,-.70) circle (1.6pt);
\fill[intrort] (0.199186,.70) circle (1.6pt);
\node[introoutline,anchor=east] at (-.29,0) {$A$};
\node[introthroat,anchor=south] at (1.70,.62) {$T$};
\draw[introthroat!65,line width=.5pt] (1.76,.64)--(1.90,.50);
\node[introbulge!85!black,anchor=south] at (2.34,.64) {$B$};
\node[intrort,anchor=north] at (1.83,-.89) {$\gamma_{A_1}$};
\draw[intrort!65,line width=.5pt] (1.83,-.86)--(1.90,-.35);
\node[align=center] at (1.40,-1.76) {(a) Small bulge:\\RT surface can enter};

\begin{scope}[shift={(5.65,0)}]
\path[fill=introoutline!7]
 (0,1.40) .. controls (.24,.88) and (.76,.70) .. (1.16,.60)
 .. controls (1.56,.50) and (1.60,.48) .. (1.90,.48)
 .. controls (2.45,.48) and (2.65,1.34) .. (3.43,1.34)
 .. controls (4.25,1.34) and (4.85,.65) .. (4.85,0)
 .. controls (4.85,-.65) and (4.25,-1.34) .. (3.43,-1.34)
 .. controls (2.65,-1.34) and (2.45,-.48) .. (1.90,-.48)
 .. controls (1.60,-.48) and (1.56,-.50) .. (1.16,-.60)
 .. controls (.76,-.70) and (.24,-.88) .. (0,-1.40) -- cycle;
\path[fill=introshadow!19]
 (1.16,.60) .. controls (1.56,.50) and (1.60,.48) .. (1.90,.48)
 .. controls (2.45,.48) and (2.65,1.34) .. (3.43,1.34)
 .. controls (4.25,1.34) and (4.85,.65) .. (4.85,0)
 .. controls (4.85,-.65) and (4.25,-1.34) .. (3.43,-1.34)
 .. controls (2.65,-1.34) and (2.45,-.48) .. (1.90,-.48)
 .. controls (1.60,-.48) and (1.56,-.50) .. (1.16,-.60)
 .. controls (1.42,-.53) and (1.42,.53) .. (1.16,.60) -- cycle;
\draw[introshadow!60,line width=.55pt]
 (1.16,-.60) .. controls (1.42,-.53) and (1.42,.53) .. (1.16,.60);
\draw[introoutline,line width=.9pt]
 (0,1.40) .. controls (.24,.88) and (.76,.70) .. (1.16,.60)
 .. controls (1.56,.50) and (1.60,.48) .. (1.90,.48)
 .. controls (2.45,.48) and (2.65,1.34) .. (3.43,1.34)
 .. controls (4.25,1.34) and (4.85,.65) .. (4.85,0)
 .. controls (4.85,-.65) and (4.25,-1.34) .. (3.43,-1.34)
 .. controls (2.65,-1.34) and (2.45,-.48) .. (1.90,-.48)
 .. controls (1.60,-.48) and (1.56,-.50) .. (1.16,-.60)
 .. controls (.76,-.70) and (.24,-.88) .. (0,-1.40);
\draw[introoutline,fill=white,line width=.9pt] (0,0) ellipse (.23 and 1.40);
\draw[introthroat,densely dashed,line width=1pt]
 (1.90,.48) arc[start angle=90,end angle=270,x radius=.10,y radius=.48];
\draw[introthroat,line width=1.3pt]
 (1.90,-.48) arc[start angle=270,end angle=450,x radius=.10,y radius=.48];
\draw[introbulge,densely dashed,line width=1pt]
 (3.43,1.34) arc[start angle=90,end angle=270,x radius=.22,y radius=1.34];
\draw[introbulge,line width=1.3pt]
 (3.43,-1.34) arc[start angle=270,end angle=450,x radius=.22,y radius=1.34];
\draw[intrort,densely dashed,line width=1.25pt]
 (-.23,0) .. controls (.10,0) and (.76,.70) .. (1.16,.60);
\draw[intrort,line width=1.5pt]
 (1.16,.60) .. controls (1.56,.50) and (.70,0) .. (.23,0);
\fill[intrort] (-.23,0) circle (1.6pt);
\fill[intrort] (.23,0) circle (1.6pt);
\node[introoutline,anchor=east] at (-.29,0) {$A$};
\node[introthroat,anchor=south] at (1.90,.57) {$T$};
\node[introbulge!85!black,anchor=south] at (3.43,1.42) {$B$};
\node[intrort,anchor=north] at (.65,-.36) {$\gamma_{A_1}$};
\draw[intrort!65,line width=.5pt] (.65,-.33)--(.64,.09);
\node[introshadow!75!black,align=center,font=\scriptsize,fill=white,
 fill opacity=.85,text opacity=1,inner sep=3pt] at (3.43,0)
 {entanglement\\shadow};
\node[align=center] at (2.43,-1.76) {(b) Large bulge:\\RT surface stays outside the shadow};
\end{scope}
\end{tikzpicture}
\endgroup
\caption{How the Python's Lunch relates to the entanglement shadow. Here $A$ is the complete boundary, $T$ the throat, $B$ the bulge, and $\gamma_{A_1}$ the RT surface for a boundary subregion $A_1\subseteq A$. (a) With a small bulge, an RT surface can pass through the throat. (b) As the bulge grows, the globally minimizing RT surface may switch to an exterior branch. The entanglement shadow (green) extends beyond the throat to the deepest reach of the RT probes.}
\label{fig:intro-single-lunch-shadow}
\end{figure}
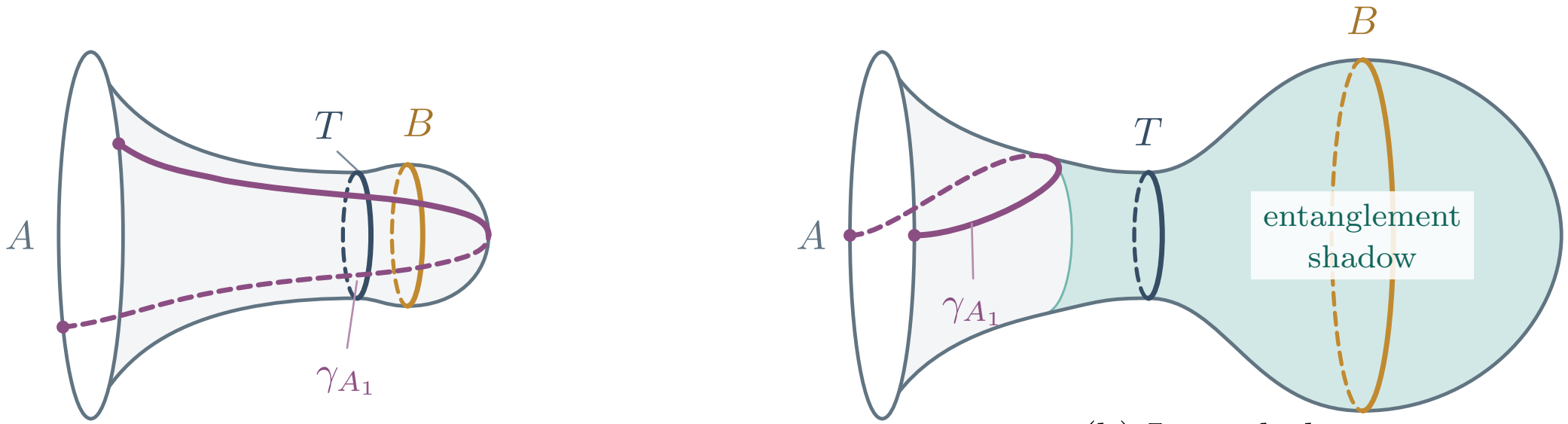

Intuitively, a larger bulge both raises the interior decoding complexity and increases the area cost for any boundary-anchored RT surface to probe deep inside (figure~\ref{fig:intro-single-lunch-shadow}). Yet a precise general relation between decoding complexity and the entanglement shadow has remained open. Here, we establish that relation by deriving upper bounds on the bulge area or generalized entropy --- and thus on the conjectured decoding complexity --- whenever an RT or QES probe enters a throat containing no smaller interior throat. The theorem is more powerful when applied to decoding a single bulk point: if its decoding complexity exceeds the bound, either the point lies inside the entanglement shadow, or the geometry realizes a special configuration with smaller throats on both sides of the probe surface. Although our cutting-and-contraction proof operates on a single Cauchy slice and leaves the fully covariant extension open, our theorems hold in both the classical RT regime and the semiclassical QES regime.

\section{Preliminaries}
\label{sec:preliminaries}

\subsection{Holographic entanglement entropy}
Let \(A\) denote the complete boundary system on a fixed bulk Cauchy slice \(\Sigma\). In the classical (\(G_N \to 0\)) limit, the Ryu--Takayanagi (RT) prescription~\cite{RyuTakayanagi2006} gives the leading-order von Neumann entanglement entropy of a boundary subregion \(A_1 \subseteq A\) as
\begin{equation}\label{eq:prelim-rt}
S(A_1) = \frac{\operatorname{Area}(\gamma_{A_1})}{4G_N},
\end{equation}
where \(\gamma_{A_1} = \partial\operatorname{EW}(A_1)\) is the codimension-two bulk surface on \(\Sigma\) that globally minimizes area among all surfaces homologous to \(A_1\). Here \(\operatorname{EW}(A_1) \subseteq \Sigma\) denotes the spatial section of the entanglement wedge of \(A_1\), whose intersection with the asymptotic boundary is \(A_1\) and whose internal bulk boundary is \(\gamma_{A_1}\).

With semiclassical bulk quantum corrections, the Engelhardt--Wall quantum extremal surface (QES) prescription~\cite{FaulknerLewkowyczMaldacena2013,EngelhardtWall2014} and its quantum maximin formulation~\cite{Wall2012,AkersEngelhardtPeningtonUsatyuk2019} select the quantum entanglement wedge \(\operatorname{EW}(A_1)\) by globally minimizing the generalized entropy over all candidate wedges homologous to \(A_1\) on the fixed slice:
\begin{equation}\label{eq:prelim-qes}
S(A_1) = S_{\rm gen}\bigl(\operatorname{EW}(A_1)\bigr) = \frac{\operatorname{Area}(\gamma_{A_1})}{4G_N} + S_{\rm bulk}\bigl(\operatorname{EW}(A_1)\bigr),
\end{equation}
where the bounding surface \(\gamma_{A_1} = \partial\operatorname{EW}(A_1)\) is the globally minimizing QES. In the quantum regime, \(\operatorname{EW}(A_1)\) may contain disconnected bulk regions --- \emph{entanglement islands} --- bounded by closed QESs whose area cost is offset by lowering \(S_{\rm bulk}\) through bulk entanglement (for instance, between black hole interior Hawking partners and an external radiation bath). 

Throughout our main theorems, we take the boundary system \(A\) to be in a pure state, so its globally minimal RT surface (or QES) is empty (\(\gamma_A = \varnothing\)) and its entanglement wedge covers the entire bulk slice (\(\operatorname{EW}(A) = \Sigma\)). In the absence of accidental degeneracy, the entanglement wedge of the complementary boundary region \(A_{\bar{1}} := A \setminus A_1\) is simply the closed complement \(\operatorname{EW}(A_1)^c := \Sigma \setminus \operatorname{Int}\operatorname{EW}(A_1)\).

In the following discussion, we also need to assign an entropy to any candidate bulk region \(D \subseteq \Sigma\). Classically, this entropy depends solely on the geometric area \(\operatorname{Area}(\partial D)\) of its internal bulk boundary \(\partial D\) (throughout, for any bulk region \(D\subseteq\Sigma\), we write \(\partial D\) for its internal bulk boundary on \(\Sigma\), excluding any segment on the asymptotic boundary). With semiclassical bulk quantum fields, we replace this classical area term on any bulk region \(D \subseteq \Sigma\) (bounded by a closed bulk hypersurface or anchored on \(A\), with no open boundary edges in the bulk) by the \emph{generalized entropy}~\cite{EngelhardtWall2014}:
\begin{equation}\label{eq:prelim-sgen}
S_{\rm gen}(D) := \frac{\operatorname{Area}(\partial D)}{4G_N} + S_{\rm bulk}(D),
\end{equation}
where \(S_{\rm bulk}(D)\) is the von Neumann entropy of the bulk fields in \(D\) (together with any fixed bath or reference subsystems), evaluated in a common ultraviolet (UV) regularization and counterterm scheme so that \(S_{\rm gen}(D)\) is UV-finite.

When the total boundary system \(A\) is in a pure state (\(\gamma_A = \varnothing\)), any closed bulk hypersurface \(X = \partial D_X\) that bounds a compact bulk region \(D_X \subset \Sigma\) disjoint from the asymptotic boundary is automatically a candidate cut homologous to the full boundary \(A\): its exterior region \(O_X := \Sigma \setminus \operatorname{Int} D_X\) (where \(\operatorname{Int} D_X\) denotes the open interior of \(D_X\)) connects \(X\) to \(A\), and \(D_X\) is the retained interior region; purity of the total state ensures \(S_{\rm gen}(D_X) = S_{\rm gen}(O_X)\). For any such cut \(X = \partial D_X\), we write \(\operatorname{Area}(X) := \operatorname{Area}(\partial D_X)\) for its classical boundary area and \(S_{\rm gen}[X] := S_{\rm gen}(D_X)\) for its generalized entropy.

\subsection{The entanglement shadow}
As one varies the boundary probe region \(A_1 \subseteq A\) over all allowed subregions, including disconnected ones, the resulting family of globally minimizing RT surfaces (or QESs) \(\{\gamma_{A_1}\}\) generally fails to foliate the entire bulk slice \(\Sigma\). Instead, competing extremal surface branches exchange dominance across first-order phase transitions, causing \(\gamma_{A_1}\) to jump discontinuously over deep bulk regions (such as the interior neighborhood of a dense star, a conical defect, or a long wormhole throat)~\cite{BalasubramanianEtAl2014,FreivogelEtAl2015}. We define the \emph{entanglement shadow} as the set of bulk points through which no globally minimizing probe surface passes,
\begin{equation}\label{eq:prelim-shadow}
\mathcal S_{\rm ent}(A) := \Sigma \setminus \bigcup_{A_1} \gamma_{A_1},
\end{equation}
and analogously \(\mathcal S_{\rm ent}^{\rm Q}(A)\) for QES probes. In both cases, all degenerate minimizing representatives are included in the union. Points inside \(\mathcal S_{\rm ent}(A)\) still lie inside the full entanglement wedge \(\operatorname{EW}(A) = \Sigma\) and can be inside \(\operatorname{EW}(A_1)\) for sufficiently large \(A_1\), but no boundary subregion's RT surface or QES ever passes through them.

\subsection{The Python's Lunch conjecture: throats, bulges, and decoding complexity}
In a Python's Lunch geometry, a boundary system or subsystem admits more than one extremal surface as an RT candidate (including the empty set $\varnothing$ when homologous). Between two locally minimal surfaces lies an expanded region containing a locally maximal (saddle) extremal surface. Because the resulting profile resembles a python after swallowing its lunch, we call the locally minimal surfaces \emph{throats} and the locally maximal saddle surfaces \emph{bulges}~\cite{BrownGharibyanPeningtonSusskind2019}.

This geometry becomes operationally sharp in the bulk-to-boundary isometry. Holographic duality acts as a quantum error-correcting code~\cite{AlmheiriDongHarlow2014} that maps bulk matter degrees of freedom (the input code space) approximately isometrically into a much larger boundary Hilbert space. This picture also applies to a boundary subsystem, say $A$: there exists an isometry from the legs cut by the RT surface $\gamma_A$ (which can be viewed as Bell pairs accounting for the entanglement entropy between $A$ and $\bar A$), together with the bulk degrees of freedom inside the entanglement wedge $\operatorname{EW}(A)$, to the output Hilbert space on the asymptotic boundary $A$. In a tensor network model of this isometry, the input legs originate from the cut along the RT surface $\gamma_A$ as well as from bulk points inside $\operatorname{EW}(A)$, while the output legs terminate on the asymptotic boundary $A$. This network is organized radially from the deep bulk interior toward the asymptotic boundary, rather than along physical time. When a locally minimal surface separates $\gamma_A$ from $A$, we call it a throat $T$, the bulk region between $T$ and $\gamma_A$ its interior (denoted $D_T$ when $\gamma_A = \varnothing$), and the region between $T$ and $A$ its outer region $O_T$.

In a Python's Lunch geometry, this tensor-network picture presents an apparent paradox: propagating from a large bulge to a smaller outer throat seems to decrease the Hilbert space dimension, blocking a naive layer-by-layer unitary or isometric map under any foliation. Nevertheless, an overall isometry from the code space to the boundary remains possible. Operationally, this mirrors a quantum computation where $a$ initial code qubits are scrambled with $b$ auxiliary qubits initialized in a fixed state and then mapped into an output Hilbert space of dimension $2^c$, with $a \le c < a+b$. Naively, the intermediate Hilbert space of dimension $2^{a+b}$ is compressed into dimension $2^c$, which cannot be achieved by an isometry on the full $2^{a+b}$-dimensional space. However, the dimension of the physical code space itself is only $2^a \le 2^c$. If we can unitarily unscramble and project out the auxiliary qubits, the overall map on the $2^a$-dimensional code space remains isometric. The most efficient known unitary algorithm to achieve this post-selection task is Grover's search algorithm~\cite{Grover1996,BrassardEtAl2000}, which requires approximately
\[
2^{\frac{a+b-c}{2}}
\]
repetitions of the scrambling and unscrambling circuit.\footnote{Reducing the Hilbert space does not necessarily imply exponential complexity. If we consider the contraction process from the AdS boundary to the bulk without a throat, this task is not exponentially complex, as we can reverse the unitary encoding circuit, which naturally unscrambles the auxiliary qubits. In a Python's Lunch, the introduction and removal of auxiliary qubits involve asymmetric circuits, leaving the post-selection with an exponentially small success probability that is hard for unitary evolution to mimic.}

From the bulk geometric perspective, a Python's Lunch illustrates the obstruction to decoding the bulk region behind $T$. We are interested in the largest cross-section inside the throat that cannot be bypassed by choosing a different foliation of the tensor network. Specifically, given a throat $T$ that separates $\gamma_A$ from the boundary $A$, the bulge $B[T,\gamma_A]$ is the minimax surface attained during an optimal sweep from $T$ to $\gamma_A$, with area
\begin{equation}
\label{eq:pl-review-width}
\operatorname{Area}\!\left(B[T,\gamma_A]\right)
:=
\min_{\{\Omega(s)\}}
\max_{s\in[0,1]}
\operatorname{Area}\!\left(\partial\Omega(s)\right),
\end{equation}
where \(\Omega(s)\) is the remaining interior region in an admissible continuous nested sweep with \(\Omega(0)=D_{T}\) and \(\Omega(1)=D_{\gamma_A}\); here \(D_{\gamma_A}:=\Sigma\setminus\operatorname{Int}\operatorname{EW}(A)\) denotes the bulk region behind \(\gamma_A\), with \(D_{\gamma_A}=\varnothing\) when \(\gamma_A=\varnothing\), and \(s\in[0,1]\) is a parameter measuring progress through the sweep. (More generally, for any two nested cuts \(X,X'\) with \(D_{X'}\subseteq D_X\), we write \(B[X,X']\) for the minimax bulge attained in an optimal sweep from \(\Omega(0)=D_X\) to \(\Omega(1)=D_{X'}\).) The RT surface, throat, and bulge may each contain multiple disconnected components.
Classically, a single lunch has decoding complexity
\begin{equation}
\label{eq:pl-review-single-classical}
\mathcal C
\sim
\mathcal C_0
\exp\!\left[
\frac{\operatorname{Area}(B)-\operatorname{Area}(T)}{8G_N}
\right].
\end{equation}

With bulk quantum corrections, each area term in \eqref{eq:pl-review-single-classical} is replaced by $4G_N$ times the generalized entropy, since the effective Hilbert space dimension of a cut receives contributions not only from its area but also from the entanglement entropy of the bulk fields (represented by bulk legs in the tensor network). From another perspective, there is a long-standing ``bag of gold'' paradox: a large bulge region can accommodate an effective bulk field-theory Hilbert space whose dimension far exceeds the finite Bekenstein--Hawking entropy bound set by the outer throat or horizon. To resolve this tension, ref.~\cite{AkersEtAl2022} proposes that the bulk-to-boundary map is not an isometry on the naive bulk effective field theory Hilbert space, but approximately preserves inner products between states of subexponential complexity. When a bulge is present, the naive bulk Hilbert space contains many null modes projected out by the holographic map.\footnote{The null states could be understood as gauge degrees of freedom, though ref.~\cite{AkersEtAl2022} does not require, or necessarily advocate, this interpretation.} In the quantum circuit picture, these redundant bulk matter modes --- together with the geometric expansion of the cross-sectional area at the bulge --- effectively act as auxiliary qubits that need to be unscrambled and post-selected via Grover's search algorithm.

Because a decoding task reconstructs bulk operators across the entire code subspace \(\mathcal H_{\rm code}\), Engelhardt, Penington, and Shahbazi-Moghaddam~\cite[section~2.1]{FindingPythons2021} propose locating the relevant quantum extremal surfaces and evaluating their generalized entropies in the maximally mixed code state \(\rho_{\rm code}=P_{\rm code}/d_{\rm code}\), where \(P_{\rm code}\) is the orthogonal projector onto \(\mathcal H_{\rm code}\) and \(d_{\rm code}=\dim\mathcal H_{\rm code}\) is its dimension. For a single bulge \(B\) and outer throat \(T\), the decoding complexity then takes the form
\begin{equation}
\label{eq:pl-review-single}
\mathcal C
\sim
\mathcal C_0
\exp\!\left[
\frac{S_{\rm gen}[B]-S_{\rm gen}[T]}{2}
\right],
\end{equation}
where \(\mathcal C_0\) is the prefactor in \eqref{eq:pl-review-single-classical}, and the generalized entropy of any cut \(X\) homologous to \(A\) is evaluated on its exterior region \(O_X\) bordering \(A\) (for a pure boundary state, this coincides with the definition \(S_{\rm gen}[X]=S_{\rm gen}(D_X)\) given above):
\begin{equation}\label{eq:pl-code-sgen}
S_{\rm gen}[X]=\frac{\operatorname{Area}(X)}{4G_N}+S_{\rm bulk}^{\rho_{\rm code}}(O_X).
\end{equation}
Here \(S_{\rm bulk}^{\rho_{\rm code}}(O_X)\) denotes the von Neumann entropy of the bulk fields in \(O_X\) in the state \(\rho_{\rm code}\). Our main theorems below rely on $A$ being in a pure state (or having negligible entropy). Nevertheless, we can always restrict \(\mathcal H_{\rm code}\) to a sufficiently small subspace --- for instance, a low-dimensional space of low-energy excitations around a background state --- so that \(S(A)=\ln d_{\rm code}\) is negligible compared to the geometric and generalized-entropy gaps of interest. In this small-code-subspace regime, the empty surface remains the globally minimizing QES of \(A\) (\(\gamma_A=\varnothing\) and \(\operatorname{EW}(A)=\Sigma\)), so \(S(A)=S_{\rm bulk}^{\rho_{\rm code}}(\Sigma)=\ln d_{\rm code}\) receives no area contribution and enters only as a small \(O(1)\) offset as \(G_N\to 0\).

More generally, a bulk geometry may contain several throats separated by bulges. For an ordered configuration, let \(T_i\) denote the throats and \(B_j\) the relevant minimax bulges, and write \(T_i\prec B_j\) when \(T_i\) lies outside \(B_j\) (between \(B_j\) and $A$). The multiple-lunch generalization~\cite{BrownGharibyanPeningtonSusskind2019,FindingPythons2021,EngelhardtPeningtonShahbaziMoghaddam2023} reads
\begin{equation}
\label{eq:pl-review-multiple}
\mathcal C
\sim
\mathcal C_0
\exp\!\left[
\frac12
\max_{T_i\prec B_j}
\left\{S_{\rm gen}[B_j]-S_{\rm gen}[T_i]\right\}
\right],
\end{equation}
or, classically,
\begin{equation}
\label{eq:pl-review-multiple-classical}
\mathcal C
\sim
\mathcal C_0
\exp\!\left[
\frac{1}{8G_N}
\max_{T_i\prec B_j}
\left\{\operatorname{Area}(B_j)-\operatorname{Area}(T_i)\right\}
\right].
\end{equation}

The maximizing pair \((T_i,B_j)\) need not be adjacent. If successive encoding segments are isometries on their respective inputs, they can be implemented sequentially so that their gate counts add and the single largest exponential cost dominates. Conversely, a further outer constriction can require nested amplitude amplification that repeatedly invokes the earlier circuit, causing the costs to multiply. Together, these two mechanisms select the largest net entropy (or area) decrease from any bulge to any subsequent outer constriction, rather than a sum of all local gaps or a comparison restricted to adjacent surfaces~\cite{BrownGharibyanPeningtonSusskind2019,FindingPythons2021}.

This ordered description assumes a compatible nested sequence of cuts. When interior throats intersect or the geometry branches, one must optimize over compatible choices of cuts and sweeps~\cite{AroraEtAl2024}.

The expressions above characterize the complexity of reconstructing the entire region down to the deepest part of the entanglement wedge. Instead, we may be interested only in decoding local operators at a specific bulk point $p\in\Sigma$. Intuitively, reconstructing $p$ should not require sweeping through deeper bottlenecks lying entirely behind an inner throat that excludes $p$. The refined point-decoding prescription~\cite[sections~2.4 and~4.1]{EngelhardtPeningtonShahbaziMoghaddam2023} therefore optimizes over admissible inner cutoffs.\footnote{We retain the Grover factor of \(1/2\) in the entropy exponent, as in eq.~(4.2) of~\cite{EngelhardtPeningtonShahbaziMoghaddam2023}; eqs.~(4.5)--(4.6) there do not display this factor explicitly.} Here a throat $T_{\rm cut}$ is called \emph{outer-minimal} if no admissible region \(D'\supseteq D_{T_{\rm cut}}\) in the same homology class has a smaller boundary area (or smaller generalized entropy in the quantum case).

\begin{definition}[Admissible Cutoffs and Point-Decoding Complexity]
\label{def:classical-point-complexity}
Let \(p\in\Sigma\) be a bulk point.
\begin{enumerate}
\item \textbf{Admissible cutoffs:} A classical throat \(T_{\rm cut}\) (possibly empty, \(T_{\rm cut}=\varnothing\)) is an \emph{admissible cutoff} for \(p\), written \(T_{\rm cut}\in\mathcal A_p\), if \(T_{\rm cut}\) is outer-minimal and \(p\) lies in the outer region \(O_{T_{\rm cut}}\), i.e., \(p\notin\operatorname{Int}D_{T_{\rm cut}}\) (in particular, \(p\) is permitted to lie on the cutoff surface \(T_{\rm cut}\) itself). Analogously, in the quantum setting, a quantum throat \(T_{\rm cut}\) (a local minimum of \(S_{\rm gen}\), possibly empty) is a \emph{quantum admissible cutoff} for \(p\), written \(T_{\rm cut}\in\mathcal A_p^{\rm Q}\), if \(T_{\rm cut}\) is outer-minimal with respect to \(S_{\rm gen}\) and \(p\notin\operatorname{Int}D_{T_{\rm cut}}\).
\item \textbf{Point-decoding complexity:} Given \(T_{\rm cut}\in\mathcal A_p\), let \(\mathcal E_{\rm out}[T_{\rm cut}]\) denote the family of classical throats \(T_0\) enclosing \(T_{\rm cut}\) (\(D_{T_{\rm cut}}\subsetneq D_{T_0}\)). We define the classical \emph{point-decoding complexity in area units} \(\Delta[p]\) of \(p\) by
\begin{equation}
\label{eq:pl-review-point}
\boxed{
\Delta[p]:=8G_N\ln\frac{\mathcal C[p]}{\mathcal C_0}
=\min_{T_{\rm cut}\in\mathcal A_p}\,\max_{T_0\in\mathcal E_{\rm out}[T_{\rm cut}]}
\Bigl\{\operatorname{Area}\!\left(B[T_0,T_{\rm cut}]\right)-\operatorname{Area}(T_0)\Bigr\},
}
\end{equation}
where the associated circuit complexity \(\mathcal C[p]\) for decoding \(p\) is conjectured to scale as
\(\mathcal C[p] \sim \mathcal C_0 \exp[\Delta[p]/(8G_N)]\).

In the quantum setting, replacing geometric area with \(4G_N S_{\rm gen}\), given \(T_{\rm cut}\in\mathcal A_p^{\rm Q}\) let \(\mathcal E_{\rm out}^{\rm Q}[T_{\rm cut}]\) denote the family of quantum throats \(T_0\) enclosing \(T_{\rm cut}\) (\(D_{T_{\rm cut}}\subsetneq D_{T_0}\)), and let \(B[T_0,T_{\rm cut}]\) be the minimax bulge of \(S_{\rm gen}\) over sweeps from \(D_{T_0}\) to \(D_{T_{\rm cut}}\). The \emph{quantum point-decoding complexity in area units} \(\Delta[p]\) is defined by
\begin{equation}
\label{eq:pl-review-point-quantum}
\boxed{
\Delta[p]:=8G_N\ln\frac{\mathcal C[p]}{\mathcal C_0}
=4G_N\min_{T_{\rm cut}\in\mathcal A_p^{\rm Q}}\,\max_{T_0\in\mathcal E_{\rm out}^{\rm Q}[T_{\rm cut}]}
\Bigl\{S_{\rm gen}\!\bigl[B[T_0,T_{\rm cut}]\bigr]-S_{\rm gen}[T_0]\Bigr\},
}
\end{equation}
with the corresponding circuit complexity
\(\mathcal C[p] \sim \mathcal C_0 \exp[\Delta[p]/(8G_N)]\).
\end{enumerate}
For an optimal cutoff \(T_{\rm cut}[p]\) (in either the classical or quantum setting), write \(T_p\) for its maximizing outer throat and \(B_p=B[T_p,T_{\rm cut}[p]]\).
\end{definition}

Unlike general decoding complexity, point-decoding complexity introduces an inner cutoff throat $T_{\rm cut}$ so the sweep can bypass the deeper interior $D_{T_{\rm cut}}$. The outer throat $T_0$ is maximized over all throats enclosing $T_{\rm cut}$, while the minimax bulge $B[T_0,T_{\rm cut}]$ only sweeps across the region between $T_0$ and $T_{\rm cut}$. Among all admissible cutoffs $T_{\rm cut}\in\mathcal A_p$, one selects the cutoff that minimizes this maximum complexity. Notably, under this definition, different bulk points share the same point-decoding complexity whenever they are not separated by any outer-minimal throat. 

This point-decoding prescription follows naturally from the bulk-to-boundary isometry picture~\cite{EngelhardtPeningtonShahbaziMoghaddam2021,EngelhardtPeningtonShahbaziMoghaddam2023}: because an outer-minimal throat $T_{\rm cut}$ has a Hilbert space dimension small enough to be mapped isometrically outward without post-selection bottlenecks, one does not need to reconstruct the circuit inside $D_{T_{\rm cut}}$ when decoding a point $p$ located outside $\operatorname{Int}D_{T_{\rm cut}}$.

The quantities above are defined on a fixed spatial slice $\Sigma$. In general time-dependent spacetimes, a covariant refinement was motivated by timelike-separated extremal surfaces that cannot be simultaneously represented on a single spatial tensor network~\cite{EngelhardtPeningtonShahbaziMoghaddam2023}; there, the bulge is selected by a \emph{maximinimax} prescription that maximizes the slice-wise minimax generalized entropy over admissible Cauchy slices. This covariant generalization is not considered in the present work.

\section{Classical case}
\label{sec:classical}

While the Python's Lunch can be defined for an arbitrary boundary subsystem, all theorems in this paper focus on the case where the decoded boundary system $A$ is in a pure state, so that its global RT surface is empty ($\gamma_{A}=\varnothing$) and $\operatorname{EW}(A)=\Sigma$. The bulk slice $\Sigma$ and boundary $A$ may have arbitrary topology and multiple connected components.

We first prove our theorems in the classical RT regime, where geometric area alone determines entanglement entropy. This setting makes the proofs relatively simple; section~\ref{sec:quantum} extends them to include bulk quantum corrections.

Throughout, we ignore accidental equalities between the areas of distinct extremal surfaces or between the area differences of distinct bulge--throat pairs. These cases are unstable under perturbation and form only a zero-measure set of all possible geometries. Because such fine-tuned coincidences only lead to unnecessary complications, we introduce the following generic assumption for the rest of the paper:
\begin{assumption}[No-Coincidence]
\label{assump:no-coincidence}
There is no accidental equality between the areas (or generalized entropies in the quantum case) of distinct extremal cuts, or between the corresponding differences of distinct bulge--throat pairs.
\end{assumption}

\subsection{Two-step sweep and fixed-throat shadow theorem}
\label{sec:classical-fixed}

Our main results rest on a two-step sweep of a throat (figure~\ref{fig:cutting-sweep}). Suppose a classical throat $T=\partial D_T$ is penetrated by the globally minimizing RT surface $\gamma_{A_1}$ of some boundary subregion $A_1\subseteq A$. We ask how large a bulge $B[T,\varnothing]$ can exist inside $D_T$. Because $\operatorname{Area}(B[T,\varnothing])$ is the minimax boundary area over all admissible sweeps contracting $D_T$ to $\varnothing$, any explicit admissible sweep from $T$ to $\varnothing$ with a known peak area immediately upper-bounds $\operatorname{Area}(B[T,\varnothing])$. We construct this sweep in two steps:

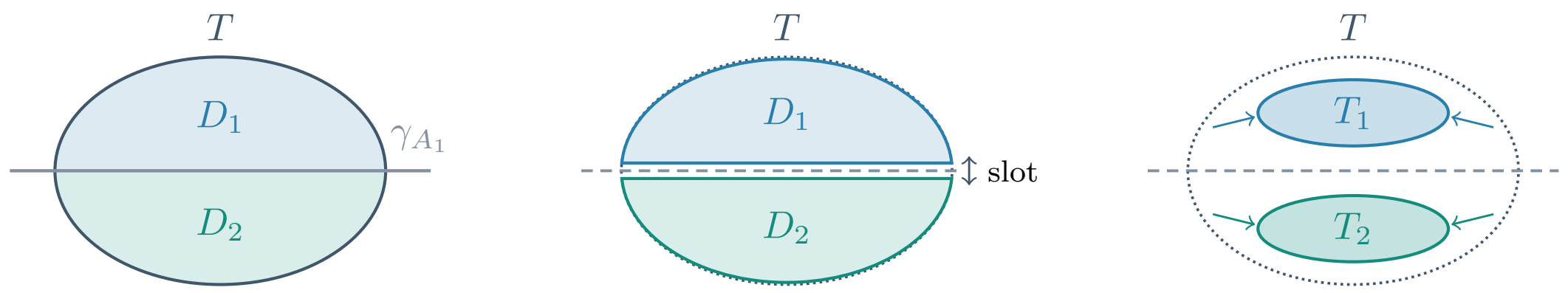
\begin{figure}[t]
\centering
\begingroup
\definecolor{fouter}{HTML}{40566B}
\definecolor{fblue}{HTML}{287EAD}
\definecolor{fgreen}{HTML}{138C7D}
\resizebox{0.98\textwidth}{!}{%
\begin{tikzpicture}[x=0.92cm,y=0.95cm,draw=fouter,text=black,
  every node/.style={font=\small}]
\begin{scope}[xshift=0cm]
\fill[fblue!16] (1.65,0) arc (0:180:1.65 and 1.10) -- cycle;
\fill[fgreen!16] (-1.65,0) arc (180:360:1.65 and 1.10) -- cycle;
\draw[line width=0.8pt] (0,0) ellipse (1.65 and 1.10);
\draw[line width=0.85pt,fouter!65] (-2.10,0)--(2.10,0);
\node[fblue] at (0,0.52) {\(D_1\)};
\node[fgreen] at (0,-0.52) {\(D_2\)};
\node[fouter,anchor=south] at (0,1.12) {\(T\)};
\node[anchor=south west,fouter!65] at (1.55,0.04) {\(\gamma_{A_1}\)};
\node at (0,-1.50) {(a) Penetrated throat \(D_T\)};
\end{scope}
\begin{scope}[xshift=5.2cm]
\draw[densely dotted,line width=0.7pt] (0,0) ellipse (1.65 and 1.10);
\fill[fblue!16] (-1.646,0.075) arc (176:4:1.65 and 1.08) -- cycle;
\fill[fgreen!16] (-1.646,-0.075) arc (184:356:1.65 and 1.08) -- cycle;
\draw[fblue,line width=0.8pt] (-1.646,0.075) arc (176:4:1.65 and 1.08) -- cycle;
\draw[fgreen,line width=0.8pt] (-1.646,-0.075) arc (184:356:1.65 and 1.08) -- cycle;
\draw[line width=0.75pt,fouter!65,densely dashed] (-2.05,0)--(1.75,0);
\draw[<->,line width=0.55pt] (1.82,-0.14)--(1.82,0.14);
\node[anchor=west,font=\scriptsize] at (1.86,0) {slot};
\node[fblue] at (0,0.55) {\(D_1\)};
\node[fgreen] at (0,-0.55) {\(D_2\)};
\node[fouter,anchor=south] at (0,1.12) {\(T\)};
\node at (0,-1.50) {(b) Cut along \(\gamma_{A_1}\cap D_T\)};
\end{scope}
\begin{scope}[xshift=10.4cm]
\draw[densely dotted,line width=0.7pt] (0,0) ellipse (1.65 and 1.10);
\draw[line width=0.75pt,fouter!65,densely dashed] (-2.05,0)--(2.05,0);
\fill[fblue!25] (0,0.56) ellipse (0.95 and 0.32);
\fill[fgreen!25] (0,-0.56) ellipse (0.95 and 0.32);
\draw[fblue,line width=0.8pt] (0,0.56) ellipse (0.95 and 0.32);
\draw[fgreen,line width=0.8pt] (0,-0.56) ellipse (0.95 and 0.32);
\draw[->,fblue,line width=0.55pt] (-1.40,0.42)--(-0.98,0.52);
\draw[->,fblue,line width=0.55pt] (1.40,0.42)--(0.98,0.52);
\draw[->,fgreen,line width=0.55pt] (-1.40,-0.42)--(-0.98,-0.52);
\draw[->,fgreen,line width=0.55pt] (1.40,-0.42)--(0.98,-0.52);
\node[fblue] at (0,0.56) {\(T_1\)};
\node[fgreen] at (0,-0.56) {\(T_2\)};
\node[fouter,anchor=south] at (0,1.12) {\(T\)};
\node at (0,-1.50) {(c) Contract to \(T_1,T_2\) (can be \(\varnothing\))};
\end{scope}
\end{tikzpicture}%
}
\endgroup
\caption{Two-step cutting and contraction sweep of a throat region \(D_T\) penetrated by an RT surface \(\gamma_{A_1}\). (a) The probe surface \(\gamma_{A_1}\) partitions \(D_T\) into \(D_1=D_T\cap\operatorname{EW}(A_1)\) and \(D_2=D_T\cap\operatorname{EW}(A_1)^c\). (b) Opening an infinitesimal slot along \(\gamma_{A_1}\cap D_T\) separates \(D_1\) and \(D_2\) with total boundary area at most \(2\operatorname{Area}(T)\). (c) Rounding the convex corners and contracting both pieces inward reduces their boundary areas monotonically until they halt at interior stopping throats \(T_1,T_2\) (which may be empty if no nontrivial throat exists).}
\label{fig:cutting-sweep}
\end{figure}

\noindent\textbf{1. Cut along the RT surface.}
First, we open an infinitesimally narrow slot through $D_T$ along a tubular neighborhood of $\gamma_{A_1}\cap D_T$. As the slot advances across a portion of $\gamma_{A_1}\cap D_T$ of area $l$, the boundary area of the remaining region increases monotonically as $\operatorname{Area}(T)+2l$, until the cut completely separates $D_T$ along $\gamma_{A_1}\cap D_T$ and the total boundary area reaches $\operatorname{Area}(T)+2\operatorname{Area}(\gamma_{A_1}\cap D_T)$. The cut splits $D_T$ into two pieces: one inside the entanglement wedge $\operatorname{EW}(A_1)$ and the other in its complement $\operatorname{EW}(A_1)^c$. Define
\begin{equation}\label{eq:classical-cut-pieces}
D_1:=D_T\cap\operatorname{EW}(A_1),\qquad
D_2:=D_T\cap\operatorname{EW}(A_1)^c.
\end{equation}
We recall the boundary-area strong subadditivity lemma used below.
\begin{lemma}[Strong Subadditivity of Boundary Area]
\label{lem:classical-submod}
For any two admissible bulk regions \(D,D'\subset\Sigma\), the boundary areas of their intersection and union satisfy
\begin{equation}\label{eq:classical-submod}
\operatorname{Area}\bigl(\partial(D\cap D')\bigr)+\operatorname{Area}\bigl(\partial(D\cup D')\bigr)
\le \operatorname{Area}(\partial D)+\operatorname{Area}(\partial D').
\end{equation}
\end{lemma}

Since \(\operatorname{EW}(A_1)\setminus D_T\) and \(\operatorname{EW}(A_1)\cup D_T\) are homologous candidate wedges for \(A_1\), global RT minimality of \(\gamma_{A_1}\) gives
\begin{equation}\label{eq:rt-slice-bound}
\operatorname{Area}(\gamma_{A_1}\cap D_T)
\le\min\Bigl\{\operatorname{Area}\bigl(T\cap\operatorname{EW}(A_1)\bigr),\,
\operatorname{Area}\bigl(T\cap\operatorname{EW}(A_1)^c\bigr)\Bigr\},
\end{equation}
and hence
\begin{equation}\label{eq:classical-piece-budget}
\operatorname{Area}(\partial D_1)\le\operatorname{Area}(T),\qquad
\operatorname{Area}(\partial D_2)\le\operatorname{Area}(T).
\end{equation}
Consequently, opening an infinitesimal slot along \(\gamma_{A_1}\cap D_T\) separates the two sides with a combined peak boundary area at most
\begin{equation}\label{eq:classical-sweep-peak}
\operatorname{Area}(\partial D_1)+\operatorname{Area}(\partial D_2)
=\operatorname{Area}(T)+2\operatorname{Area}(\gamma_{A_1}\cap D_T)
\le 2\operatorname{Area}(T).
\end{equation}

\noindent\textbf{2. Contract the two pieces.}
Next, each of the pieces $D_1$ and $D_2$ is bounded by segments of two locally minimal surfaces ($T$ and $\gamma_{A_1}$). Consequently, each piece is locally outward-minimizing and its boundary area cannot decrease under local outward deformations, whereas at points where $\gamma_{A_1}$ intersects $T$ transversely, a convex corner is formed. Rounding these convex corners inward guarantees an initial contraction with strictly decreasing boundary area. Continuing this area-nonincreasing inward contraction shrinks each piece $D_i$ ($i=1,2$) monotonically until it halts at a locally minimal stopping throat \(T_i=\partial D_{T_i}\) (which may be empty, $T_i=\varnothing$). Since \(\gamma_{A_1}\) enters \(\operatorname{Int}D_T\), each nonempty \(D_{T_i}\subseteq D_i\subsetneq D_T\) is bounded by a throat distinct from \(T\), with strictly smaller area \(\operatorname{Area}(T_i)<\operatorname{Area}(T)\).

Now, if no nonempty smaller throat exists inside $D_T$, the inward contractions of $D_1$ and $D_2$ cannot stop until both pieces have shrunk all the way to the empty set $\varnothing$. This explicitly constructs a complete sweep from $T$ to $\varnothing$ whose peak boundary area is $\operatorname{Area}(T)+2\operatorname{Area}(\gamma_{A_1}\cap D_T)\le 2\operatorname{Area}(T)$.

Conversely, if there exists a smaller throat inside $D_1$ or $D_2$, the monotonically decreasing sweep halts at that interior throat, and sweeping across its interior requires the boundary area to increase again; the overall peak area then depends on the deeper geometry inside that smaller throat.

This two-step sweep yields our first conditional shadow theorem:
\begin{theorem}[Classical Fixed-Throat Entanglement Shadow]
\label{thm:cl-fixed}
Let \(T\) be a classical throat in the pure-state bulk slice \(\Sigma\). Suppose there is no nonempty classical throat \(T'\) inside \(D_{T}\) with \(\operatorname{Area}(T')<\operatorname{Area}(T)\). Then the minimax bulge $B=B[T,\varnothing]$ inside $T$ satisfies
\begin{equation}\label{eq:cl-fixed-threshold}
\boxed{\quad
\operatorname{Area}\!\left(B[T,\varnothing]\right)>2\operatorname{Area}(T)
\quad\Longrightarrow\quad
\operatorname{Int}D_{T}\subseteq\mathcal S_{\rm ent}(A).
\quad}
\end{equation}
\end{theorem}

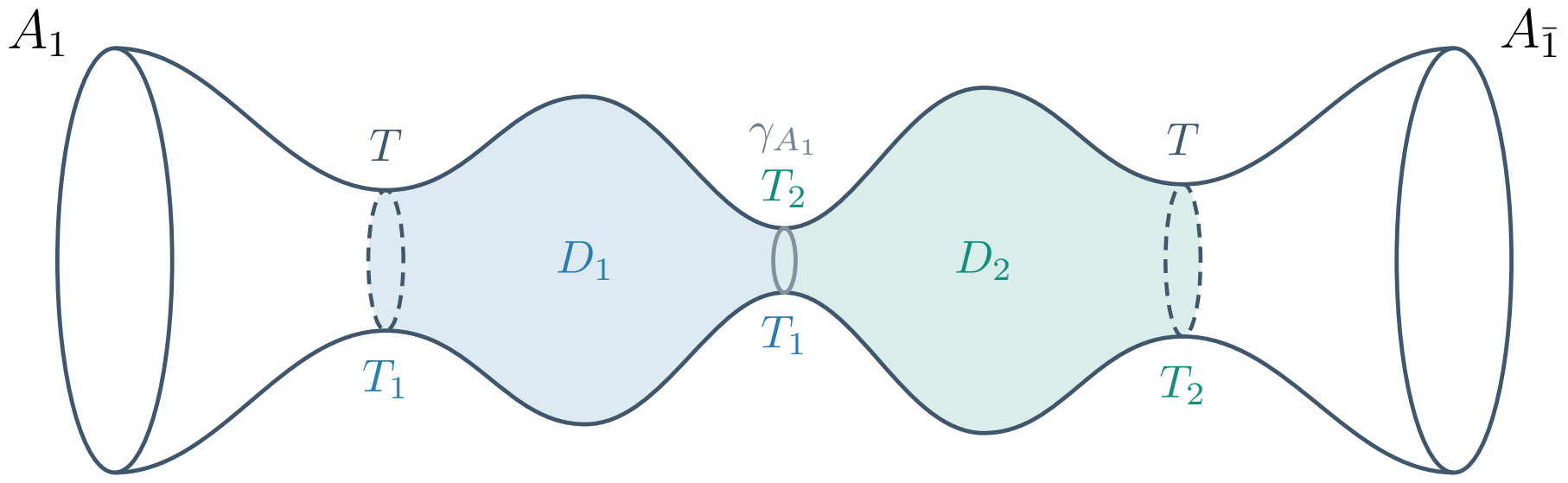
\begin{figure}[t]
\centering
\begingroup
\definecolor{fouter}{HTML}{40566B}
\definecolor{fblue}{HTML}{287EAD}
\definecolor{fgreen}{HTML}{138C7D}
\begin{tikzpicture}[x=1.22cm,y=1.12cm,draw=fouter,text=black,
  every node/.style={font=\small}]
\path[use as bounding box] (-5.05,-1.85) rectangle (5.05,1.95);
\fill[fblue!16]
  (-2.50,0.48)
  .. controls (-1.95,0.48) and (-1.75,1.12) .. (-1.25,1.12)
  .. controls (-0.75,1.12) and (-0.45,0.22) .. (0.00,0.22)
  -- (0.00,-0.22)
  .. controls (-0.45,-0.22) and (-0.75,-1.12) .. (-1.25,-1.12)
  .. controls (-1.75,-1.12) and (-1.95,-0.48) .. (-2.50,-0.48)
  -- cycle;
\fill[fgreen!16]
  (0.00,0.22)
  .. controls (0.45,0.22) and (0.75,1.18) .. (1.25,1.18)
  .. controls (1.75,1.18) and (1.95,0.52) .. (2.50,0.52)
  -- (2.50,-0.52)
  .. controls (1.95,-0.52) and (1.75,-1.18) .. (1.25,-1.18)
  .. controls (0.75,-1.18) and (0.45,-0.22) .. (0.00,-0.22)
  -- cycle;
\fill[fblue!16] (-2.50,0) ellipse (0.11 and 0.48);
\fill[fgreen!16] (2.50,0) ellipse (0.11 and 0.52);
\draw[line width=0.9pt]
  (-4.20,1.45)
  .. controls (-3.55,1.45) and (-3.15,0.48) .. (-2.50,0.48)
  .. controls (-1.95,0.48) and (-1.75,1.12) .. (-1.25,1.12)
  .. controls (-0.75,1.12) and (-0.45,0.22) .. (0.00,0.22)
  .. controls (0.45,0.22) and (0.75,1.18) .. (1.25,1.18)
  .. controls (1.75,1.18) and (1.95,0.52) .. (2.50,0.52)
  .. controls (3.15,0.52) and (3.55,1.45) .. (4.20,1.45);
\draw[line width=0.9pt]
  (-4.20,-1.45)
  .. controls (-3.55,-1.45) and (-3.15,-0.48) .. (-2.50,-0.48)
  .. controls (-1.95,-0.48) and (-1.75,-1.12) .. (-1.25,-1.12)
  .. controls (-0.75,-1.12) and (-0.45,-0.22) .. (0.00,-0.22)
  .. controls (0.45,-0.22) and (0.75,-1.18) .. (1.25,-1.18)
  .. controls (1.75,-1.18) and (1.95,-0.52) .. (2.50,-0.52)
  .. controls (3.15,-0.52) and (3.55,-1.45) .. (4.20,-1.45);
\draw[line width=0.9pt] (-4.20,0) ellipse (0.36 and 1.45);
\draw[line width=0.9pt] (4.20,0) ellipse (0.36 and 1.45);
\draw[densely dashed,line width=0.85pt] (-2.50,0) ellipse (0.11 and 0.48);
\draw[fouter!65,line width=0.9pt] (0.00,0) ellipse (0.07 and 0.22);
\draw[densely dashed,line width=0.85pt] (2.50,0) ellipse (0.11 and 0.52);
\node[font=\large] at (-4.68,1.55) {\(A_1\)};
\node[font=\large] at (4.68,1.55) {\(A_{\bar{1}}\)};
\node[fouter,anchor=south] at (-2.50,0.56) {\(T\)};
\node[fblue,anchor=north] at (-2.50,-0.56) {\(T_{1}\)};
\node[fouter!75,anchor=south] at (0.00,0.60) {\(\gamma_{A_1}\)};
\node[fgreen,anchor=south] at (0.00,0.24) {\(T_{2}\)};
\node[fblue,anchor=north] at (0.00,-0.26) {\(T_{1}\)};
\node[fouter,anchor=south] at (2.50,0.60) {\(T\)};
\node[fgreen,anchor=north] at (2.50,-0.60) {\(T_{2}\)};
\node[fblue] at (-1.25,0) {\(D_1\)};
\node[fgreen] at (1.25,0) {\(D_2\)};
\end{tikzpicture}
\endgroup
\caption{Cutting a two-sided wormhole throat region \(D_T\) (bounded by the two outer throat components labeled \(T\)) along a closed RT surface \(\gamma_{A_1}\). After cutting along \(\gamma_{A_1}\), the boundaries of the two resulting pieces \(D_1\) and \(D_2\) are locally minimal throats \(T_1=\partial D_1\) and \(T_2=\partial D_2\) whose boundary components coincide with the components of \(T\) and \(\gamma_{A_1}\). In this case, the contraction step terminates immediately.}
\label{fig:wormhole-closed-rt}
\end{figure}

\phantomsection\label{rem:classical-closed-rt}
This theorem applies to arbitrary spatial dimensions and boundary topologies. Importantly, a globally minimizing RT surface $\gamma_{A_1}$ that enters $\operatorname{Int}D_T$ need not intersect the throat $T$. Instead, $\gamma_{A_1}$ can be an isolated closed minimal surface component inside $\operatorname{Int}D_T$ that does not touch $T$. In that case, after cutting $D_T$ along $\gamma_{A_1}$ into $D_1$ and $D_2$, the second contraction step terminates immediately because the boundaries $T_{1}:=\partial D_1$ and $T_{2}:=\partial D_2$ have no corners and already form locally minimal throats whose surface components coincide with $\gamma_{A_1}$ and the components of $T$. Therefore, Theorem~\ref{thm:cl-fixed} still applies directly to such isolated RT surfaces. For example, this situation naturally occurs when considering a Python's Lunch inside a two-sided wormhole connecting $A_1$ and $A_{\bar{1}}$, where $T$ consists of the two outer constrictions near the two boundaries and $\gamma_{A_1}$ is the closed minimal neck in the middle; cutting $D_T$ along $\gamma_{A_1}$ leaves two regions $D_1,D_2$ bounded by the two locally minimal throats $T_{1}$ (consisting of the left component of $T$ together with $\gamma_{A_1}$) and $T_{2}$ (consisting of $\gamma_{A_1}$ together with the right component of $T$), as illustrated in figure~\ref{fig:wormhole-closed-rt}.

The hypothesis of Theorem~\ref{thm:cl-fixed} --- that no smaller throat exists inside $D_T$ --- is restrictive, but in geometries with only a single throat and a single bulge (as studied in~\cite{BrownGharibyanPeningtonSusskind2019}), it directly yields a bound on the interior bulge area $\operatorname{Area}(B[T,\varnothing])$ (and hence on the resulting decoding complexity) outside the entanglement shadow. Similarly, when the bulk contains a nested sequence of throats ($\cdots\subset D_{T_2}\subset D_{T_1}$), one can construct a sweep from $T_1$ to $T_2$ to bound the intermediate bulge area $\operatorname{Area}(B[T_1,T_2])$ whenever an RT surface penetrates $T_1$ without entering $T_2$.

Theorem~\ref{thm:cl-fixed} requires an unconditional absence of smaller interior throats because such throats halt the contraction. Since the RT surface does not penetrate those stopping throats, we ask whether a sharper bound holds for the point-dependent decoding complexity $\Delta[p]$ in \eqref{eq:pl-review-point}, whose inner cutoff $T_{\rm cut}$ bypasses deep bottlenecks. Based on the point-dependent decoding complexity, we establish a general theorem showing that an RT surface can pass through a high-complexity point $p$ only in a specific configuration.

\subsection{Point-dependent decoding complexity}
\label{sec:classical-point}

We now consider the point-dependent complexity in Definition~\ref{def:classical-point-complexity}. Previous work left the relation between $p$ and $T_p$ unconstrained; we first show the intuitive result that \(T_p\) is always outer-minimal and encloses the point $p$.

\begin{lemma}[Target Point Lies Inside the Optimal Outer Throat]
\label{lem:cl-p-interior}
Suppose \(\Delta[p]>0\). Then the maximizing outer throat \(T_p\) in Definition~\ref{def:classical-point-complexity} is outer-minimal and satisfies
\begin{equation}\label{eq:p-in-int-Tp}
p\in\operatorname{Int}D_{T_p}.
\end{equation}
\end{lemma}

\begin{proof}
First, suppose \(T_p\) were not outer-minimal, and let \(T_E\) be a least-area cut enclosing \(D_{T_p}\), with \(\operatorname{Area}(T_E)<\operatorname{Area}(T_p)\).\footnote{We assume the least-area enclosure exists with a regular throat boundary and that the intersection sweeps below are admissible after smoothing corners.} For any sweep \(\Omega(s)\) from \(T_E\) to \(T_{\rm cut}[p]\), the restricted sweep \(\Omega(s)\cap D_{T_p}\) sweeps from \(T_p\) to \(T_{\rm cut}[p]\), while \(\Omega(s)\cup D_{T_p}\) still encloses \(D_{T_p}\) and thus has boundary area at least \(\operatorname{Area}(T_E)\). Applying strong subadditivity of area (Lemma~\ref{lem:classical-submod}) to \(\Omega(s)\) and \(D_{T_p}\) therefore gives
\[
\operatorname{Area}\!\left(\partial(\Omega(s)\cap D_{T_p})\right)-\operatorname{Area}(T_p)
\le \operatorname{Area}(\partial\Omega(s))-\operatorname{Area}(T_E).
\]
Since \(\Delta[p]\) is the infimum of \(\max_s\operatorname{Area}(\partial\Omega'(s))-\operatorname{Area}(T_p)\) over all admissible sweeps \(\Omega'(s)\) from \(D_{T_p}\) to \(D_{T_{\rm cut}[p]}\), taking the infimum over \(\Omega(s)\) gives \(\Delta[p]\le\operatorname{Area}(B[T_E,T_{\rm cut}[p]])-\operatorname{Area}(T_E)\), contradicting the uniqueness of \(T_p\) as the maximizer of the bulge--throat gap (Assumption~\ref{assump:no-coincidence}). Thus \(T_p\) is outer-minimal.

Second, suppose \(p\notin\operatorname{Int}D_{T_p}\), so \(T_p\in\mathcal A_p\) is itself an admissible cutoff. By minimality of \(T_{\rm cut}[p]\) in \eqref{eq:pl-review-point}, there is a throat \(T_0\in\mathcal E_{\rm out}[T_p]\) (outer-minimal by the same argument as above) with maximal \(\operatorname{Area}(B[T_0,T_p])-\operatorname{Area}(T_0)\ge\Delta[p]>0\). Since \(T_0\in\mathcal E_{\rm out}[T_{\rm cut}[p]]\) and \(T_0\neq T_p\), \(\operatorname{Area}(B[T_0,T_{\rm cut}[p]])-\operatorname{Area}(T_0)<\Delta[p]\) under Assumption~\ref{assump:no-coincidence}. Therefore, there exists a sweep \(\Omega(s)\) from \(D_{T_0}\) to \(D_{T_{\rm cut}[p]}\) whose peak satisfies
\[
\max_{s\in[0,1]}\operatorname{Area}\bigl(\partial\Omega(s)\bigr)
<\operatorname{Area}(T_0)+\Delta[p]
\le\operatorname{Area}\!\left(B[T_0,T_p]\right).
\]
Consider \(\Omega(s)\cap D_{T_p}\) and \(\Omega(s)\cup D_{T_p}\):
\begin{itemize}
\item If \(\operatorname{Area}(\partial(\Omega(s)\cap D_{T_p}))\ge\operatorname{Area}(T_p)\) for all \(s\in[0,1]\), Lemma~\ref{lem:classical-submod} keeps \(\operatorname{Area}(\partial(\Omega(s)\cup D_{T_p}))\le\max_s\operatorname{Area}(\partial\Omega(s))<\operatorname{Area}(B[T_0,T_p])\), giving a sweep from \(D_{T_0}\) to \(D_{T_p}\) below \(\operatorname{Area}(B[T_0,T_p])\), a contradiction.
\item Otherwise, \(\operatorname{Area}(\partial(\Omega(s)\cap D_{T_p}))\) drops below \(\operatorname{Area}(T_p)\). Before the first drop, the restricted sweep must attain a boundary area of at least \(\operatorname{Area}(B[T_p,T_{\rm cut}[p]])\): otherwise, there is a smaller outer throat with at least the same minimax bulge area, contradicting the maximality of \(T_p\). Hence there exists a parameter \(s_*\) prior to the first drop such that
\[
\operatorname{Area}\bigl(\partial(\Omega(s_*)\cap D_{T_p})\bigr)
\ge \operatorname{Area}\!\left(B[T_p,T_{\rm cut}[p]]\right)
= \operatorname{Area}(T_p)+\Delta[p].
\]
For all \(s\in[0,s_*]\), we have \(\operatorname{Area}(\partial(\Omega(s)\cap D_{T_p}))\ge\operatorname{Area}(T_p)\), so Lemma~\ref{lem:classical-submod} keeps the union sweep \(\Omega(s)\cup D_{T_p}\) strictly below \(\operatorname{Area}(B[T_0,T_p])\), and at \(s=s_*\) gives
\[\begin{aligned}
\operatorname{Area}\bigl(\partial(\Omega(s_*)\cup D_{T_p})\bigr)
&< \operatorname{Area}(T_0)+\Delta[p]+\operatorname{Area}(T_p)-\bigl(\operatorname{Area}(T_p)+\Delta[p]\bigr) \\
&= \operatorname{Area}(T_0).
\end{aligned}\]
This yields another throat in \(\mathcal E_{\rm out}[T_p]\) with area strictly smaller than \(\operatorname{Area}(T_0)\) and minimax bulge area at least \(\operatorname{Area}(B[T_0,T_p])\), contradicting the choice of \(T_0\).
\end{itemize}
Hence \(p\in\operatorname{Int}D_{T_p}\).
\end{proof}

Now we can state the point-decoding shadow theorem:
\begin{theorem}[Classical Point-Decoding Entanglement Shadow Theorem]
\label{thm:cl-point}
Suppose a bulk point \(p\in\Sigma\) satisfies
\begin{equation}\label{eq:cl-point-threshold}
\Delta[p]>\operatorname{Area}(T_p).
\end{equation}
If \(p\) lies on a globally minimal RT surface \(\gamma_{A_1}\) for some subsystem \(A_1\subseteq A\) (whether \(\gamma_{A_1}\) is boundary-anchored or contains closed components not reaching the asymptotic boundary), then there must exist two nonempty stopping throats \(T_1\) and \(T_2\), whose retained regions satisfy
\begin{equation}\label{eq:cl-two-sides-location}
D_{T_1}\subseteq D_{T_p}\cap\operatorname{EW}(A_1),\qquad
D_{T_2}\subseteq D_{T_p}\cap\operatorname{EW}(A_1)^c,
\end{equation}
such that:
\begin{enumerate}
\item Each \(T_i\) (\(i=1,2\)) is an admissible cutoff in \(\mathcal A_p\), but \(D_{T_1}\cup D_{T_2}\) cannot be enclosed by any single admissible cutoff in \(\mathcal A_p\).
\item Their areas obey the strict individual and joint bounds
\begin{equation}\label{eq:cl-two-sided-area-bounds}
0<\operatorname{Area}(T_i)<\operatorname{Area}(T_p)\quad(i=1,2),\qquad
\operatorname{Area}(T_1)+\operatorname{Area}(T_2)>\operatorname{Area}(T_p).
\end{equation}
\end{enumerate}
Contrapositively, if \(D_{T_p}\) contains no such ``throats-on-both-sides'' configuration, namely \(T_1,T_2\in\mathcal A_p\) satisfying \eqref{eq:cl-two-sides-location} and \eqref{eq:cl-two-sided-area-bounds} with no common admissible cutoff in \(\mathcal A_p\), then \(\Delta[p]>\operatorname{Area}(T_p)\) implies that no globally minimizing RT surface \(\gamma_{A_1}\) can pass through \(p\), hence \(p\in\mathcal S_{\rm ent}(A)\).
\end{theorem}

\begin{figure}[t]
\centering
\begingroup
\definecolor{sboutline}{HTML}{617482}
\definecolor{sbthroat}{HTML}{354D65}
\definecolor{sbcut}{HTML}{148779}
\definecolor{sbbulge}{HTML}{C18A30}
\definecolor{sbrt}{HTML}{8B4E83}
\begin{tikzpicture}[scale=0.74,x=0.93cm,y=0.93cm,
  every node/.style={font=\small},line cap=round,line join=round]
\path[fill=sboutline!8,draw=sboutline,line width=0.95pt]
  (-4.30,5.40)
  .. controls (-2.15,4.87) and (-1.06,3.35) .. (-1.06,2.10)
  .. controls (-1.13,1.17) and (-1.57,0.80) .. (-2.03,0.28)
  .. controls (-2.33,-0.12) and (-3.43,-0.60) .. (-3.52,-1.72)
  .. controls (-3.64,-2.94) and (-2.76,-3.54) .. (-1.90,-3.54)
  .. controls (-0.72,-3.54) and (-0.58,-2.45) .. (-0.65,-1.60)
  .. controls (-0.73,-0.84) and (-0.99,-0.42) .. (-0.87,-0.04)
  .. controls (-0.72,0.42) and (-0.39,0.65) .. (0.03,0.67)
  .. controls (0.40,0.66) and (0.63,0.34) .. (0.72,-0.05)
  .. controls (0.93,-0.74) and (0.57,-1.39) .. (1.12,-1.98)
  .. controls (1.69,-2.56) and (2.76,-2.18) .. (2.93,-1.49)
  .. controls (3.16,-0.61) and (2.42,-0.16) .. (2.08,0.29)
  .. controls (1.58,0.85) and (1.13,1.19) .. (1.06,2.10)
  .. controls (1.06,3.35) and (2.15,4.87) .. (4.30,5.40)
  arc[start angle=0,end angle=180,x radius=4.30,y radius=1.00]
  -- cycle;
\draw[fill=white,draw=sboutline!85,line width=0.95pt]
  (0,5.40) ellipse (4.30 and 1.00);
\draw[sbrt,densely dashed,line width=1.15pt]
  (0,6.40) .. controls (-0.18,5.20) and (-0.18,3.30) .. (-0.16,2.40)
  .. controls (-0.13,1.58) and (-0.25,1.18) .. (0.03,0.67);
\draw[sbrt,line width=1.4pt]
  (0.03,0.67) .. controls (0.65,0.79) and (0.62,1.64) .. (0.44,2.24)
  .. controls (0.29,2.94) and (0.13,3.75) .. (0,4.40);
\draw[sbthroat,densely dashed,line width=1.15pt]
  (1.06,2.10) arc[start angle=0,end angle=180,x radius=1.06,y radius=0.31];
\draw[sbthroat,line width=1.55pt]
  (-1.06,2.10) arc[start angle=180,end angle=360,x radius=1.06,y radius=0.31];
\begin{scope}[shift={(-1.45,0.12)},rotate=-15.5]
\draw[sbcut,densely dashed,line width=1.2pt]
  (0.60,0) arc[start angle=0,end angle=180,x radius=0.60,y radius=0.22];
\draw[sbcut,line width=1.7pt]
  (-0.60,0) arc[start angle=180,end angle=360,x radius=0.60,y radius=0.22];
\end{scope}
\begin{scope}[shift={(1.40,0.12)},rotate=14]
\draw[sbthroat,densely dashed,line width=1.1pt]
  (0.70,0) arc[start angle=0,end angle=180,x radius=0.70,y radius=0.22];
\draw[sbthroat,line width=1.45pt]
  (-0.70,0) arc[start angle=180,end angle=360,x radius=0.70,y radius=0.22];
\end{scope}
\draw[sbbulge,densely dashed,line width=1.3pt]
  (-1.62,0.82) .. controls (-0.95,1.55) and (1.13,1.30) .. (2.12,0.12)
  .. controls (2.60,-0.40) and (2.92,-0.70) .. (2.98,-1.18);
\draw[sbbulge,line width=1.8pt]
  (-1.62,0.82) .. controls (-0.98,0.55) and (-0.17,1.20) .. (0.46,0.93)
  .. controls (0.78,0.60) and (0.88,-0.53) .. (0.85,-1.18)
  arc[start angle=180,end angle=360,x radius=1.065,y radius=0.32];
\fill[sbrt] (0,6.40) circle (1.6pt);
\fill[sbrt] (0,4.40) circle (1.6pt);
\fill[white] (0.03,0.67) circle (3.1pt);
\fill[sbrt] (0.03,0.67) circle (2.1pt);
\node[sbrt,anchor=east] at (-0.12,0.72) {$p$};
\node[sbthroat,anchor=south] at (3.15,6.30) {boundary $A$};
\node[sbrt,anchor=west] at (0.93,3.48) {$\gamma_{A_1}$ (RT)};
\draw[sbrt!70,line width=0.5pt] (0.80,3.45)--(0.22,3.35);
\node[sbthroat,anchor=west,align=left] at (2.45,2.06) {$T_p$\\outer throat};
\draw[sbthroat!65,line width=0.55pt] (2.29,2.04)--(1.10,2.10);
\node[sbcut,anchor=east,align=right] at (-2.90,0.75) {$T_1=T_{\rm cut}[p]$\\cutoff};
\draw[sbcut!75,line width=0.55pt] (-2.73,0.66)--(-1.86,0.29);
\node[sbthroat,anchor=west,align=left] at (2.95,0.70) {$T_2$\\throat};
\draw[sbthroat!65,line width=0.55pt] (2.78,0.58)--(2.05,0.28);
\node[sbbulge!90!black,anchor=west,align=left] at (3.48,-1.18) {$B_p$\\bulge};
\draw[sbbulge!80,line width=0.55pt] (3.33,-1.18)--(3.02,-1.18);
\node[text=sbthroat,align=center] at (0,-4.18)
  {$B_p=B[T_p,T_{\rm cut}[p]]$};
\end{tikzpicture}
\endgroup
\caption{Illustration of the throats-on-both-sides exception in Theorem~\ref{thm:cl-point}. The RT surface $\gamma_{A_1}$ passes through $p$ between the inner throats $T_1,T_2$, with $T_{\rm cut}[p]=T_1$ as the illustrated cutoff. The gold closed curve denotes the bulge $B_p=B[T_p,T_{\rm cut}[p]]$: the minimax surface of the sweep between the cutoff and the outer throat $T_p$, which determines the complexity of decoding the point $p$. This two-sided obstruction permits an arbitrarily large decoding complexity while the RT surface passes through $p$.}
\label{fig:cl-unbounded-two-sided}
\end{figure}
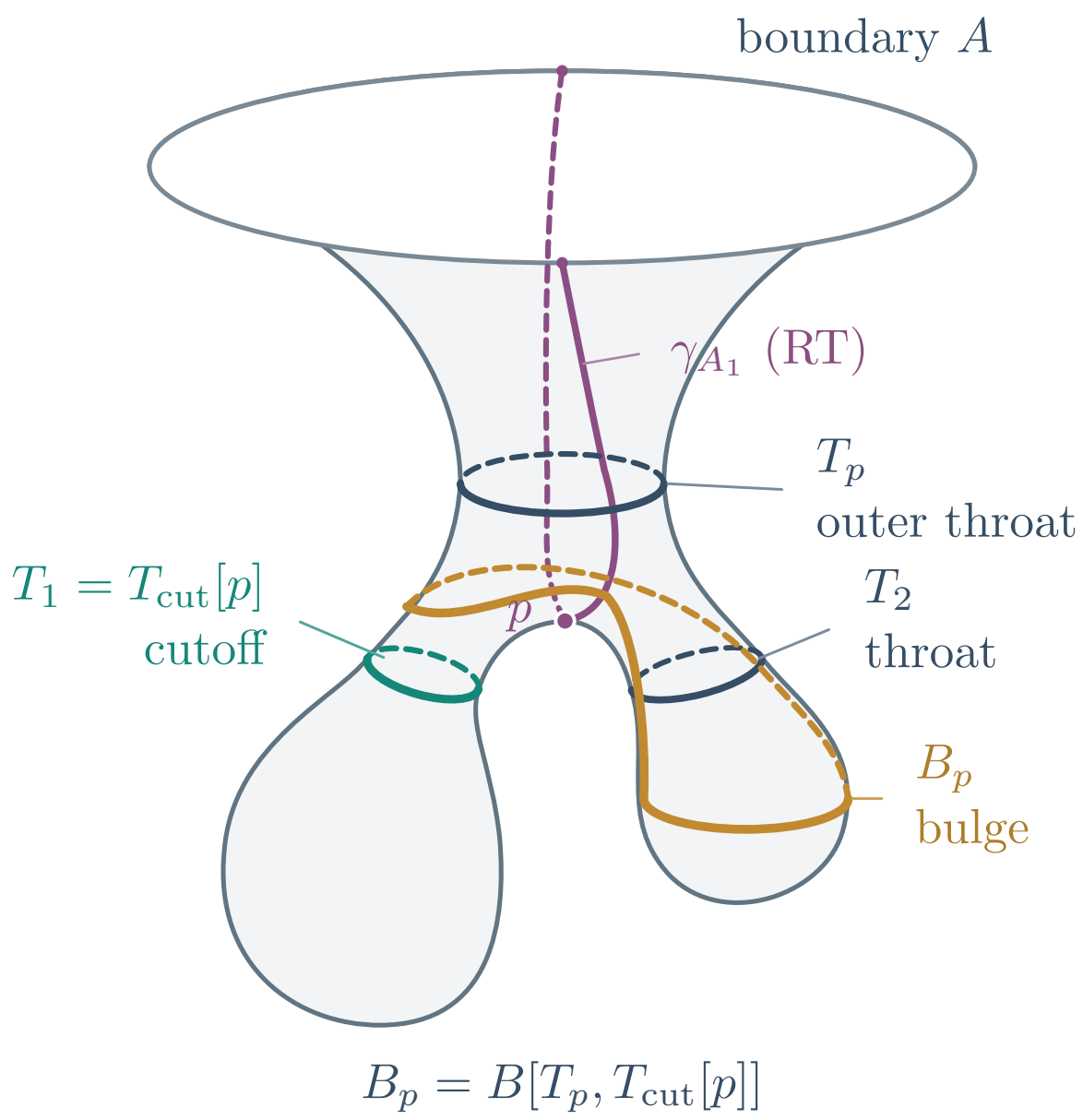

Figure~\ref{fig:cl-unbounded-two-sided} illustrates this throats-on-both-sides exception, in which $\Delta[p]$ can grow arbitrarily large while $\gamma_{A_1}$ still passes through $p$. Before proving Theorem~\ref{thm:cl-point}, we note two consequences of the theorem. First, if we consider a geometry where all outer-minimal throats form a single nested chain, two nonoverlapping admissible cutoffs on opposite sides of $\gamma_{A_1}$ cannot exist:
\begin{corollary}[Shadow Criterion for Nested Throats]
\label{cor:cl-single-chain}
Suppose the bulk slice \(\Sigma\) has a single nested chain of outer-minimal throats (for instance, when the relevant throat class consists of concentric spheres), so that for any two outer-minimal throats \(T_1,T_2\), one encloses the other (\(D_{T_1}\subseteq D_{T_2}\) or \(D_{T_2}\subseteq D_{T_1}\)). Then the throats-on-both-sides exception of Theorem~\ref{thm:cl-point} cannot occur, and
\begin{equation}\label{eq:cl-single-chain-shadow}
\Delta[p]>\operatorname{Area}(T_p)
\quad\Longrightarrow\quad
p\in\mathcal S_{\rm ent}(A).
\end{equation}
\end{corollary}

Second, the point-selected outer throat $T_p$ can be difficult to locate explicitly in a complicated geometry, whereas the outermost throat $T_{\rm outermost}$ enclosing $T_p$ can be easier to identify. We therefore obtain:
\begin{corollary}[Outer Enclosing Throat Threshold]
\label{cor:cl-outermost}
Let \(T_{\rm outermost}\) denote the outermost classical throat enclosing \(D_{T_p}\). Because \(T_p\) is outer-minimal (Lemma~\ref{lem:cl-p-interior}), we have \(\operatorname{Area}(T_p)\le\operatorname{Area}(T_{\rm outermost})\). Therefore, whenever the throats-on-both-sides exception of Theorem~\ref{thm:cl-point} is absent,
\begin{equation}\label{eq:cl-outermost-shadow}
\Delta[p]>\operatorname{Area}(T_{\rm outermost})
\quad\Longrightarrow\quad
p\in\mathcal S_{\rm ent}(A),
\end{equation}
providing a practical sufficient condition when the exact point-dependent throat $T_p$ is difficult
to locate explicitly. 

Engelhardt and Wall~\cite{EngelhardtWall2017,EngelhardtWall2018} identify the area of such an outermost surface (which may or may not be the globally minimal surface), in units of \(4G_N\), with the ``simple entropy'' of the boundary system: a coarse-grained entropy that retains only the information accessible to simple boundary experiments.  

\end{corollary}

To prove Theorem~\ref{thm:cl-point}, we first show that any optimal cutoff $T_{\rm cut}$ can be replaced by a least-area enclosure of its cut-and-shrunk remnants (continuing to disregard accidental exact area equalities between distinct extremal cuts or distinct bulge--throat pairs).
\begin{lemma}[Least-Area Remnant Enclosure as the Optimal Cutoff]
\label{lem:cl-cutoff-replace}
Suppose \(\Delta[p]>\operatorname{Area}(T_p)\) and \(p\in\gamma_{A_1}\) for a globally minimal RT surface \(\gamma_{A_1}\). Let \(T_{\rm cut}\in\mathcal A_p\) be an optimal cutoff for \(\Delta[p]\). Cut \(D_{T_{\rm cut}}\) along \(\gamma_{A_1}\), shrink both sides by the cutting-and-contraction argument in the proof of Theorem~\ref{thm:cl-fixed}, and let \(D_R\) denote the union of the resulting stable remnant regions (if \(\gamma_{A_1}\) does not enter \(\operatorname{Int}D_{T_{\rm cut}}\), then \(D_R=D_{T_{\rm cut}}\)). Let \(\widehat T_{\rm cut}=\partial D_{\widehat T_{\rm cut}}\) be a least-area cut enclosing \(D_R\) chosen inside \(D_{T_{\rm cut}}\). Then:
\begin{enumerate}
\item \(\widehat T_{\rm cut}\) is an admissible cutoff in \(\mathcal A_p\) satisfying
\begin{equation}\label{eq:cl-hat-cutoff-nesting}
D_R\subseteq D_{\widehat T_{\rm cut}}\subseteq D_{T_{\rm cut}},\qquad
\operatorname{Area}(\widehat T_{\rm cut})\le\operatorname{Area}(T_{\rm cut})\le\operatorname{Area}(T_p)<\Delta[p].
\end{equation}
\item \(\widehat T_{\rm cut}\) is also an optimal cutoff for \(\Delta[p]\) and shares the same dominant bulge--throat pair \((B_p,T_p)\). In particular, we may assume without loss of generality that the optimal cutoff \(T_{\rm cut}\) is already a least-area enclosure of its cut-and-shrunk remnants \(D_R\).
\end{enumerate}
\end{lemma}

\begin{figure}[t]
\centering
\begingroup
\definecolor{fouter}{HTML}{40566B}
\definecolor{fblue}{HTML}{287EAD}
\definecolor{fgreen}{HTML}{138C7D}
\definecolor{fpurple}{HTML}{8153B5}
\definecolor{forange}{HTML}{DE981F}
\definecolor{fred}{HTML}{C64352}
\begin{tikzpicture}[x=1.6cm,y=1.15cm,line cap=round,
 every node/.style={font=\small}]
\path[use as bounding box] (-3.2,-2.45) rectangle (3.2,2.15);
\filldraw[fill=fpurple!5,draw=fpurple,line width=1.3pt]
 (.15,0) ellipse (2.65 and 1.85);
\filldraw[fill=fouter!5,draw=fouter,dashed,line width=1pt]
 (0,0) ellipse (2.25 and 1.5);
\filldraw[fill=forange!20,draw=forange,line width=1pt]
 (-.75,0) ellipse (1.05 and 1.1);
\fill[fblue,fill opacity=.10] (.75,0) ellipse (1.6 and 1.2);
\draw[fblue,line width=1pt] (.75,0) ellipse (1.6 and 1.2);
\filldraw[fill=fblue!20,draw=fblue,line width=1.1pt]
 (-.1,0) ellipse (.32 and .43);
\foreach \y in {.19,-.19}{
 \filldraw[fill=fgreen!40,draw=fgreen,line width=.8pt]
 (-.1,\y) ellipse (.075 and .105);}
\draw[fouter!45,line width=.7pt] (-2.95,0)--(2.95,0);
\fill[fred] (.53,0) circle (.045);
\node[fred,anchor=south west] at (.53,.04) {$p$};
\node[forange!75!black] at (-1.15,.77) {$T_{\rm cut}$};
\node[fblue] at (1.28,.85) {$T_X$};
\node[fpurple,anchor=south] at (.15,1.92) {$T_E$};
\node[fouter,anchor=west] at (1.6,-1.65) {$T_p$ (dashed)};
\draw[fouter,->] (1.59,-1.58)--(1.18,-1.28);
\node[fblue,anchor=east] at (-1.65,-1.55) {$\widehat T_{\rm cut}$};
\draw[fblue,->] (-1.6,-1.5)--(-.34,-.29);
\node[fgreen] at (-.15,.75) {$D_R$};
\draw[fgreen,->] (-.13,.61)--(-.1,.30);
\node[fouter!75,anchor=south east] at (2.95,.05) {$\gamma_{A_1}$};
\node at (0,-2.15)
 {$D_R\subseteq D_{\widehat T_{\rm cut}}\subseteq D_{T_{\rm cut}}\subseteq D_{T_p}$};
\end{tikzpicture}
\endgroup
\caption{Illustration of Lemma~\ref{lem:cl-cutoff-replace}. The dashed slate curve is the original maximizing outer throat $T_p$; the orange curve is $T_{\rm cut}$. The green remnants $D_R$ lie in their shaded blue least-area enclosure $D_{\widehat T_{\rm cut}}$. The blue exterior candidate $T_X$ may cross $T_{\rm cut}$. The purple $T_E$ is a least-area enclosure of $D_{T_X}\cup D_{T_{\rm cut}}$ and may extend outside $T_p$. The red point $p$ lies on the gray RT surface $\gamma_{A_1}$.}
\label{fig:two-sided-hull}
\end{figure}
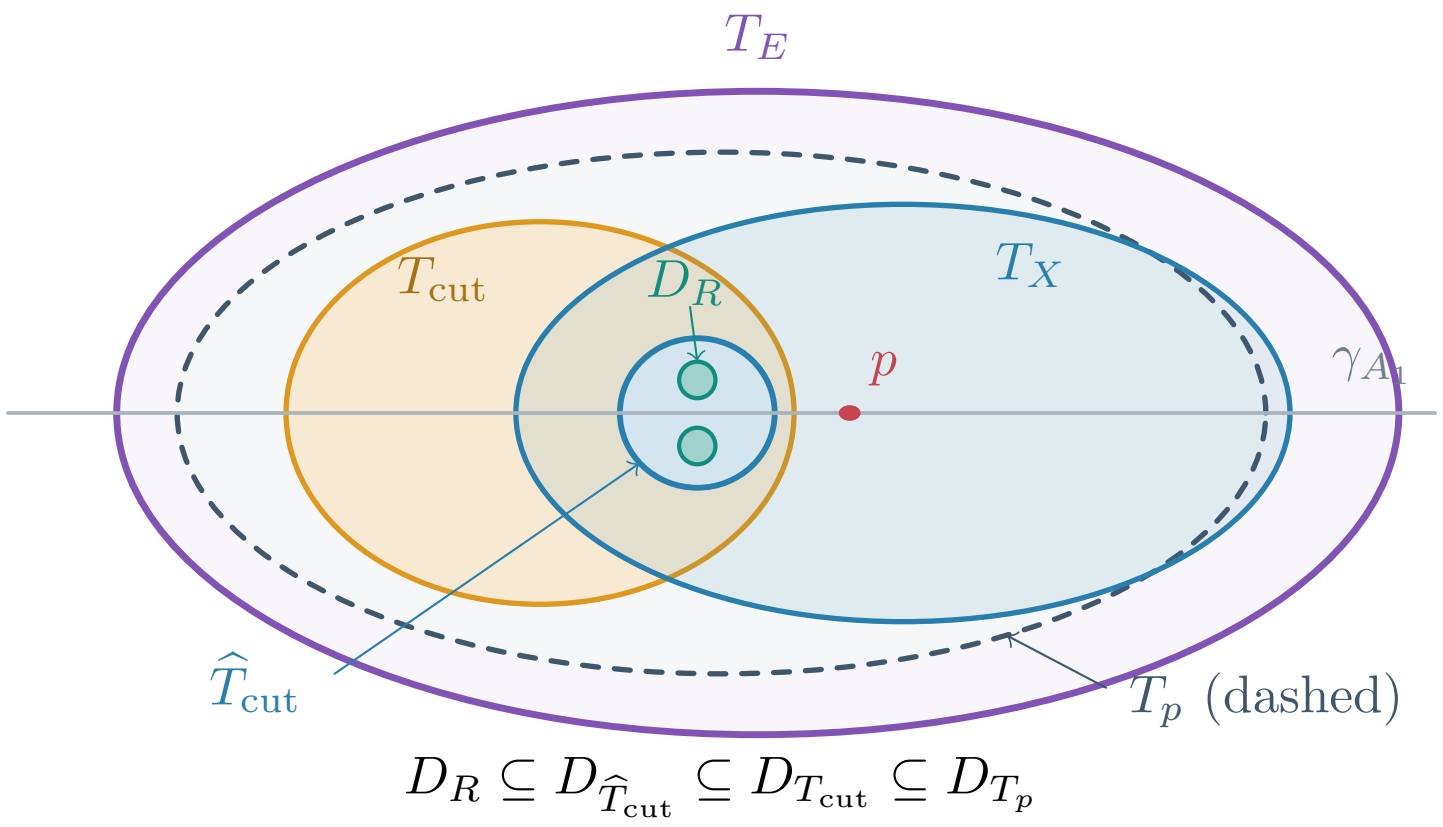

\begin{proof}
For the original optimal cutoff \(T_{\rm cut}\), the assumption \(\Delta[p]>\operatorname{Area}(T_p)\) gives
\begin{equation}\label{eq:two-sided-main-width}
\operatorname{Area}\!\left(B[T_p,T_{\rm cut}]\right)
=\operatorname{Area}(T_p)+\Delta[p]
>2\operatorname{Area}(T_p).
\end{equation}
Since \(T_{\rm cut}\) is outer-minimal and enclosed by \(T_p\), we have \(\operatorname{Area}(T_{\rm cut})\le\operatorname{Area}(T_p)<\Delta[p]\). Moreover, because \(T_p\) maximizes the complexity over all exterior candidates \(T_X\in\mathcal E_{\rm out}[T_{\rm cut}]\), every throat \(T_X\) enclosing \(T_{\rm cut}\) obeys
\begin{equation}\label{eq:two-sided-old-max}
\operatorname{Area}\!\left(B[T_X,T_{\rm cut}]\right)-\operatorname{Area}(T_X)\le\Delta[p].
\end{equation}

Because \(D_R\) is bounded by locally minimal throats and \(D_{T_{\rm cut}}\) is outer-minimal, a least-area region enclosing \(D_R\) is itself an outer-minimal throat region and can be chosen inside \(D_{T_{\rm cut}}\), so \(D_{\widehat T_{\rm cut}}\subseteq D_{T_{\rm cut}}\) and \(\operatorname{Area}(\widehat T_{\rm cut})\le\operatorname{Area}(T_{\rm cut})\). Because \(p\notin\operatorname{Int}D_{T_{\rm cut}}\), we also have \(p\notin\operatorname{Int}D_{\widehat T_{\rm cut}}\), and therefore \(\widehat T_{\rm cut}\in\mathcal A_p\) (choosing \(\widehat T_{\rm cut}=\varnothing\) if \(D_R=\varnothing\)).

To prove that \(\widehat T_{\rm cut}\) is also an optimal cutoff, it suffices (by the outer-minimality of maximizing outer throats shown in Lemma~\ref{lem:cl-p-interior}) to consider an arbitrary outer-minimal exterior candidate \(T_X\in\mathcal E_{\rm out}[\widehat T_{\rm cut}]\); then
\begin{equation}\label{eq:two-sided-nesting}
D_R\subseteq D_{\widehat T_{\rm cut}}\subseteq D_{T_X}\cap D_{T_{\rm cut}}\subseteq D_{T_{\rm cut}}.
\end{equation}
Let \(T_E\) be a least-area cut enclosing \(D_{T_X}\cup D_{T_{\rm cut}}\) (see figure~\ref{fig:two-sided-hull}). Since \(D_{T_X}\) and \(D_{T_{\rm cut}}\) are throat regions, \(T_E\) is itself a throat enclosing \(T_{\rm cut}\), so \eqref{eq:two-sided-old-max} provides a sweep \(\{\Omega(s)\}\) from \(T_E\) to \(T_{\rm cut}\) with peak area at most \(\operatorname{Area}(T_E)+\Delta[p]\). We now construct a two-part sweep from \(T_X\) to \(\widehat T_{\rm cut}\):

\emph{Part 1 (\(T_X\to\partial(D_{T_X}\cap D_{T_{\rm cut}})\)):} Intersect each retained region \(\Omega(s)\) of the \(T_E\to T_{\rm cut}\) sweep with \(D_{T_X}\) to form \(\Omega'(s)=\Omega(s)\cap D_{T_X}\). Since \(\Omega(s)\cup D_{T_X}\) encloses \(D_{T_{\rm cut}}\cup D_{T_X}\), the definition of \(T_E\) implies \(\operatorname{Area}(\partial(\Omega(s)\cup D_{T_X}))\ge\operatorname{Area}(T_E)\). Lemma~\ref{lem:classical-submod} then gives
\[
\operatorname{Area}\bigl(\partial(\Omega(s)\cap D_{T_X})\bigr)
\le\operatorname{Area}\bigl(\partial\Omega(s)\bigr)+\operatorname{Area}(T_X)-\operatorname{Area}\bigl(\partial(\Omega(s)\cup D_{T_X})\bigr)
\le\operatorname{Area}(T_X)+\Delta[p].
\]

\emph{Part 2 (\(\partial(D_{T_X}\cap D_{T_{\rm cut}})\to\widehat T_{\rm cut}\)):} Since both \(T_X\) and \(T_{\rm cut}\) are outer-minimal and \(D_{T_X}\cup D_{T_{\rm cut}}\) encloses each of them, Lemma~\ref{lem:classical-submod} yields
\begin{equation}\label{eq:two-sided-j-bound}
\operatorname{Area}\bigl(\partial(D_{T_X}\cap D_{T_{\rm cut}})\bigr)
\le\min\bigl\{\operatorname{Area}(T_X),\operatorname{Area}(T_{\rm cut})\bigr\}.
\end{equation}
By \eqref{eq:two-sided-nesting}, intersecting the contractions of the two halves of \(D_{T_{\rm cut}}\) with \(D_{T_X}\cap D_{T_{\rm cut}}\) contracts the two cut halves of \(D_{T_X}\cap D_{T_{\rm cut}}\) monotonically to the same remnants \(D_R\), with peak area at most \(2\operatorname{Area}(\partial(D_{T_X}\cap D_{T_{\rm cut}}))\) by the two-side budget \eqref{eq:classical-sweep-peak}. Taking the union of each stage of this contraction with \(D_{\widehat T_{\rm cut}}\) does not increase boundary area (by Lemma~\ref{lem:classical-submod}, since the intersection with \(D_{\widehat T_{\rm cut}}\) still encloses \(D_R\) and \(\widehat T_{\rm cut}\) is a least-area enclosure of \(D_R\)). This produces a sweep \(\partial(D_{T_X}\cap D_{T_{\rm cut}})\to\widehat T_{\rm cut}\) with peak area at most \(2\operatorname{Area}(\partial(D_{T_X}\cap D_{T_{\rm cut}}))\).

Joining Parts~1 and 2 and applying \eqref{eq:two-sided-j-bound} and \eqref{eq:cl-hat-cutoff-nesting}, we find
\begin{equation}\label{eq:two-sided-replacement}
\begin{aligned}
\operatorname{Area}\!\left(B[T_X,\widehat T_{\rm cut}]\right)-\operatorname{Area}(T_X)
&\le\max\Bigl\{\Delta[p],\,2\operatorname{Area}\bigl(\partial(D_{T_X}\cap D_{T_{\rm cut}})\bigr)-\operatorname{Area}(T_X)\Bigr\}\\
&\le\max\bigl\{\Delta[p],\operatorname{Area}(T_{\rm cut})\bigr\}
=\Delta[p].
\end{aligned}
\end{equation}
Thus the maximum complexity over all exterior candidates \(T_X\) for \(\widehat T_{\rm cut}\) is at most \(\Delta[p]\). Because \(\Delta[p]\) is already the minimum over \(\mathcal A_p\), equality must hold, so \(\widehat T_{\rm cut}\) is optimal. Assumption~\ref{assump:no-coincidence} then identifies its dominant pair with \((B_p,T_p)\).
\end{proof}

We now prove Theorem~\ref{thm:cl-point}.

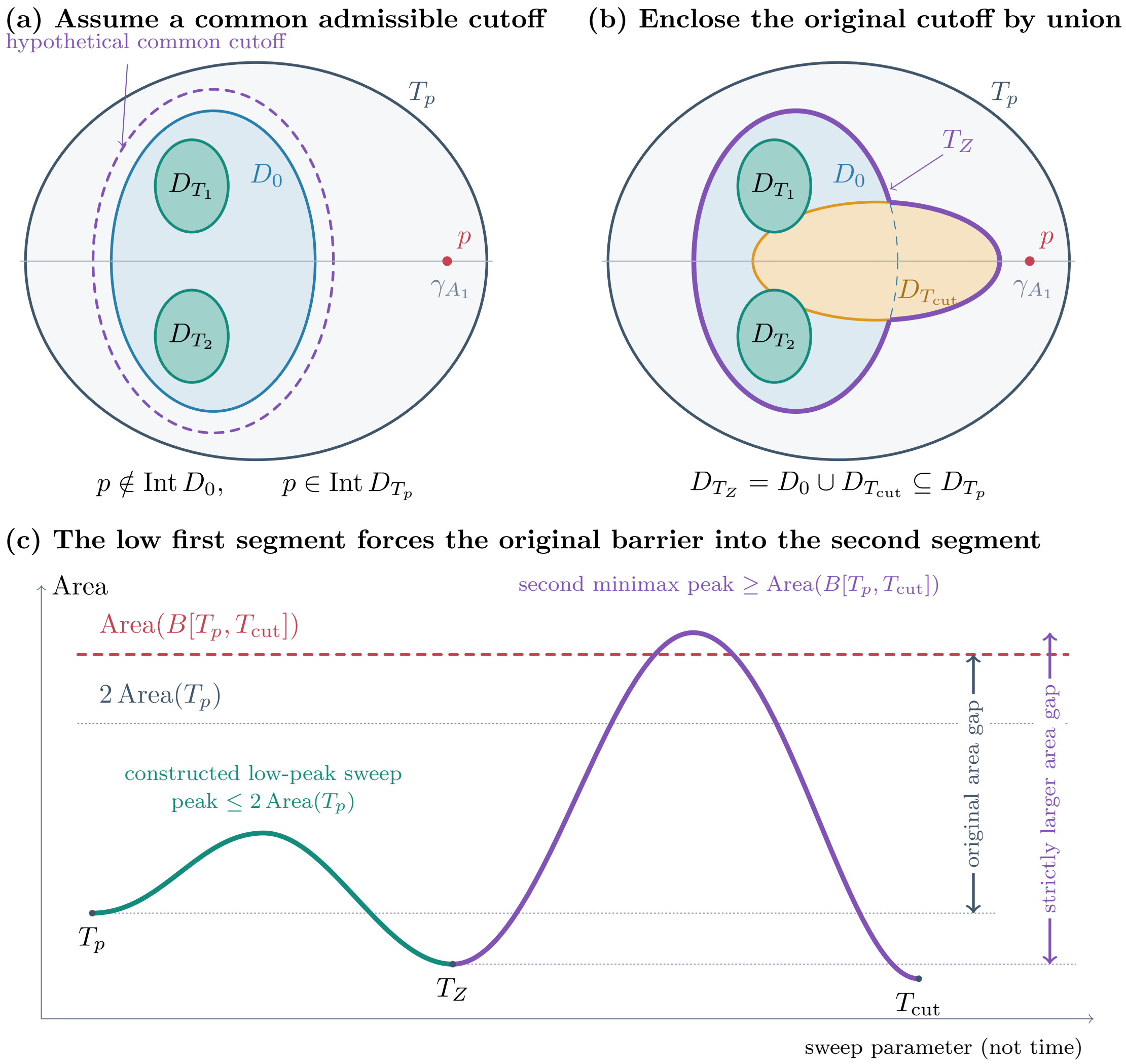
\begin{figure}[p]
\centering
\begingroup
\definecolor{fouter}{HTML}{40566B}
\definecolor{fblue}{HTML}{287EAD}
\definecolor{fgreen}{HTML}{138C7D}
\definecolor{fpurple}{HTML}{8153B5}
\definecolor{forange}{HTML}{DE981F}
\definecolor{fred}{HTML}{C64352}
\begin{tikzpicture}[x=0.95cm,y=0.95cm,every node/.style={font=\small},line cap=round]
\path[use as bounding box] (0,-.25) rectangle (15.6,14.25);
\node[anchor=west,font=\small\bfseries] at (0,14) {(a) Assume a common admissible cutoff};
\node[anchor=west,font=\small\bfseries] at (8,14) {(b) Enclose the original cutoff by union};
\begin{scope}[shift={(3.6,10.7)},x=1.4cm,y=1.4cm]
\filldraw[fill=fouter!5,draw=fouter,line width=1pt] (0,0) ellipse (2.15 and 1.85);
\draw[fpurple,dashed,line width=1pt] (-.4,0) ellipse (1.12 and 1.6);
\filldraw[fill=fblue!16,draw=fblue,line width=1pt] (-.4,0) ellipse (.95 and 1.4);
\foreach \y/\i in {.7/1,-.7/2}{
 \filldraw[fill=fgreen!40,draw=fgreen,line width=1pt] (-.6,\y) ellipse (.34 and .43);
 \node at (-.6,\y) {$D_{T_\i}$};}
\draw[fouter!45] (-2.2,0)--(2.2,0);
\fill[fred] (1.78,0) circle (.045);
\node[fred,anchor=south west] at (1.78,.02) {$p$};
\node[fouter] at (1.55,1.55) {$T_p$};
\node[fblue] at (.1,0.8) {$D_0$};
\node[fouter!75,anchor=north] at (1.82,-.08) {$\gamma_{A_1}$};
\node[fpurple,anchor=south,font=\scriptsize] at (-1.03,1.85) {hypothetical common cutoff};
\draw[fpurple,->] (-1.2,1.83)--(-1.25,1.07);
\end{scope}
\node at (3.6,7.62) {$p\notin\operatorname{Int}D_0,\qquad p\in\operatorname{Int}D_{T_p}$};
\begin{scope}[shift={(11.6,10.7)},x=1.4cm,y=1.4cm]
\filldraw[fill=fouter!5,draw=fouter,line width=1pt] (0,0) ellipse (2.15 and 1.85);
\fill[fblue!16] (-.4,0) ellipse (.95 and 1.4);
\filldraw[fill=forange!27,draw=forange,line width=1pt] (.35,0) ellipse (1.15 and .55);
\draw[fblue,dashed] (-.4,0) ellipse (.95 and 1.4);
\begin{scope}
\pgfseteorule\clip (-3,-3) rectangle (3,3) (.35,0) ellipse (1.15 and .55);
\draw[fpurple,line width=1.8pt] (-.4,0) ellipse (.95 and 1.4);
\end{scope}
\begin{scope}
\pgfseteorule\clip (-3,-3) rectangle (3,3) (-.4,0) ellipse (.95 and 1.4);
\draw[fpurple,line width=1.8pt] (.35,0) ellipse (1.15 and .55);
\end{scope}
\foreach \y/\i in {.7/1,-.7/2}{
 \filldraw[fill=fgreen!40,draw=fgreen,line width=1pt] (-.6,\y) ellipse (.34 and .43);
 \node at (-.6,\y) {$D_{T_\i}$};}
\draw[fouter!45] (-2.2,0)--(2.2,0);
\fill[fred] (1.78,0) circle (.045);
\node[fred,anchor=south west] at (1.78,.02) {$p$};
\node[fouter] at (1.55,1.55) {$T_p$};
\node[fblue] at (.1,0.8) {$D_0$};
\node[forange!75!black] at (.84,-.29) {$D_{T_{\rm cut}}$};
\node[fouter!75,anchor=north] at (1.82,-.08) {$\gamma_{A_1}$};
\node[fpurple] at (1.12,1.12) {$T_Z$};
\draw[fpurple,->] (.97,.99)--(.48,.67);
\end{scope}
\node at (11.6,7.62) {$D_{T_Z}=D_0\cup D_{T_{\rm cut}}\subseteq D_{T_p}$};
\node[anchor=west,font=\small\bfseries] at (0,6.85) {(c) The low first segment forces the original barrier into the second segment};
\draw[->,fouter!80] (.65,.3)--(.65,6.25) node[right,text=black] {$\operatorname{Area}$};
\draw[->,fouter!80] (.65,.3)--(15.1,.3);
\node[anchor=north east,font=\scriptsize] at (15.1,.18) {sweep parameter (not time)};
\draw[fred,dashed,line width=1pt] (1.15,5.3)--(14.75,5.3);
\node[fred,anchor=south west] at (1.3,5.35) {$\operatorname{Area}(B[T_p,T_{\rm cut}])$};
\draw[fouter!65,densely dotted] (1.15,4.35)--(14.75,4.35);
\node[fouter,anchor=south west] at (1.3,4.39) {$2\operatorname{Area}(T_p)$};
\draw[fouter!65,densely dotted] (1.35,1.75)--(13.75,1.75);
\draw[fpurple!70,densely dotted] (6.3,1.05)--(14.85,1.05);
\draw[fgreen,line width=1.8pt] (1.35,1.75)
 .. controls (2.3,1.75) and (2.8,2.85) .. (3.7,2.85)
 .. controls (4.6,2.85) and (5.35,1.05) .. (6.3,1.05);
\draw[fpurple,line width=1.8pt] (6.3,1.05)
 .. controls (7.55,1.05) and (8.45,5.6) .. (9.6,5.6)
 .. controls (10.85,5.6) and (11.7,.85) .. (12.7,.85);
\node[fgreen,align=center,font=\scriptsize] at (3.7,3.45) {constructed low-peak sweep\\peak $\le 2\operatorname{Area}(T_p)$};
\node[fpurple,anchor=south,font=\scriptsize] at (10.1,5.95)
 {second minimax peak $\ge \operatorname{Area}(B[T_p,T_{\rm cut}])$};
\foreach \x/\y/\lab in {1.35/1.75/T_p,6.3/1.05/T_Z,12.7/.85/T_{\rm cut}}{
 \fill[fouter] (\x,\y) circle (.045);
 \node[anchor=north] at (\x,\y-.05) {$\lab$};}
\draw[<->,fouter,line width=1pt] (13.45,1.75)--(13.45,5.3)
 node[midway,fill=white,inner sep=2pt,rotate=90,font=\scriptsize] {original area gap};
\draw[<->,fpurple,line width=1pt] (14.5,1.05)--(14.5,5.6)
 node[midway,fill=white,inner sep=2pt,rotate=90,font=\scriptsize] {strictly larger area gap};
\end{tikzpicture}
\endgroup
\caption{Why a common admissible cutoff leads to a contradiction in the proof of Theorem~\ref{thm:cl-point}. (a) Such a cutoff would contain a least-area enclosure $D_0$ of the two stopping regions. Since $p\notin\operatorname{Int}D_0$ but $p\in\operatorname{Int}D_{T_p}$, the assumed nondegeneracy gives a strictly smaller boundary area for $D_0$. (b) Strong subadditivity and the least-area enclosure property make $T_Z=\partial(D_0\cup D_{T_{\rm cut}})$ an exterior throat candidate with $\operatorname{Area}(T_Z)<\operatorname{Area}(T_p)$. It need not itself be an admissible inner cutoff excluding $p$. (c) Taking the union of the constructed sweep with $D_{T_Z}$ gives a first segment with peak area at most $2\operatorname{Area}(T_p)$, below the original bulge. Concatenation therefore forces $\operatorname{Area}(B[T_Z,T_{\rm cut}])\ge \operatorname{Area}(B[T_p,T_{\rm cut}])$. The lower starting throat gives a strictly larger bulge--throat area gap, contradicting the maximizing property of $T_p$. The nontrivial case $T_Z\ne T_{\rm cut}$ is shown; equality already contradicts the original minimax height. Regions and profiles are schematic, not a metric realization or numerical solution; the second profile represents the minimax height comparison.}
\label{fig:two-sided-cutoff}
\end{figure}

\begin{proof}[Proof of Theorem~\ref{thm:cl-point}]
\medskip\noindent\textbf{1. Construct the two side stopping throats.}
By Lemma~\ref{lem:cl-p-interior}, \(p\in\operatorname{Int}D_{T_p}\), and by Lemma~\ref{lem:cl-cutoff-replace} we may choose the optimal cutoff \(T_{\rm cut}\in\mathcal A_p\) to be a least-area enclosure of its remnants \(D_R\). Cut \(D_{T_p}\) along \(\gamma_{A_1}\) and contract both sides to stopping throats \(T_1,T_2\), either of which may initially be empty (figure~\ref{fig:two-sided-cutoff}).

Because \(D_{T_{\rm cut}}\subset D_{T_p}\), each side piece of \(D_{T_p}\) contains the corresponding side remnant of \(D_{T_{\rm cut}}\) in \(D_R\). We can choose the contractions of the two halves of \(D_{T_p}\) to retain the already stable remnants \(D_R\), ensuring \(D_R\subseteq D_{T_1}\cup D_{T_2}\). The cutting-and-contraction argument of Theorem~\ref{thm:cl-fixed}, applied to \(D_T=D_{T_p}\), directly supplies the sweep \(T_p\to\partial(D_{T_1}\cup D_{T_2})\) with peak area at most \(2\operatorname{Area}(T_p)\) and the strict bound \(\operatorname{Area}(T_i)<\operatorname{Area}(T_p)\) for each nonempty \(T_i\).

Furthermore, because \(T_p\) is outer-minimal (Lemma~\ref{lem:cl-p-interior}) and \(\gamma_{A_1}\) is globally minimal, each cut piece of \(D_{T_p}\) is globally outward-minimizing, and this outward minimality is preserved under inward contraction, making each nonempty stopping throat \(T_i\) outer-minimal. Finally, since \(D_{T_1}\subseteq\operatorname{EW}(A_1)\) and \(D_{T_2}\subseteq\operatorname{EW}(A_1)^c\) while \(p\in\gamma_{A_1}=\partial\operatorname{EW}(A_1)\), \(p\) cannot lie in the open interior \(\operatorname{Int}D_{T_i}\) of either side (whether \(\gamma_{A_1}\) is boundary-anchored or a closed RT component), so \(T_i\in\mathcal A_p\).

\medskip\noindent\textbf{2. Rule out a low-area enclosure of both remnants.}
Suppose for contradiction that there exists an enclosure of \(D_{T_1}\cup D_{T_2}\) inside \(D_{T_p}\) with boundary area strictly smaller than \(\operatorname{Area}(T_p)\). As illustrated in figure~\ref{fig:two-sided-cutoff}(a), since \(T_p\) is outer-minimal, we can choose a least-area enclosure \(D_0\subseteq D_{T_p}\) of \(D_{T_1}\cup D_{T_2}\) inside \(D_{T_p}\); then \(\partial D_0\) is an outer-minimal throat with \(\operatorname{Area}(\partial D_0)<\operatorname{Area}(T_p)\).

Because \(D_R\subseteq D_{T_1}\cup D_{T_2}\subseteq D_0\) and \(D_R\subseteq D_{T_{\rm cut}}\), the intersection \(D_0\cap D_{T_{\rm cut}}\) still encloses \(D_R\), as illustrated in figure~\ref{fig:two-sided-cutoff}(b). Since \(T_{\rm cut}\) is a least-area enclosure of \(D_R\), we have \(\operatorname{Area}(\partial(D_0\cap D_{T_{\rm cut}}))\ge\operatorname{Area}(T_{\rm cut})\). Applying strong subadditivity (Lemma~\ref{lem:classical-submod}) to \(D_0\) and \(D_{T_{\rm cut}}\) then gives
\[
\operatorname{Area}\bigl(\partial(D_0\cup D_{T_{\rm cut}})\bigr)
\le\operatorname{Area}(\partial D_0)<\operatorname{Area}(T_p).
\]
Since \(D_0\cup D_{T_{\rm cut}}\subseteq D_{T_p}\) is itself an enclosure of \(D_{T_1}\cup D_{T_2}\) and \(D_0\) already minimizes boundary area among all such enclosures, \(D_0\cup D_{T_{\rm cut}}\) is also a least-area enclosure of \(D_{T_1}\cup D_{T_2}\). Hence its boundary
\[
T_Z:=\partial(D_0\cup D_{T_{\rm cut}})
\]
is a genuine throat enclosing \(T_{\rm cut}\), with
\begin{equation}\label{eq:two-sided-z-small}
\operatorname{Area}(T_Z)<\operatorname{Area}(T_p).
\end{equation}

As illustrated in figure~\ref{fig:two-sided-cutoff}(c), take the sweep \(T_p\to\partial(D_{T_1}\cup D_{T_2})\) from Step~1 and form the union of every stage of that sweep with \(D_0\cup D_{T_{\rm cut}}\). Because the intersection of each stage with \(D_0\cup D_{T_{\rm cut}}\) still encloses \(D_{T_1}\cup D_{T_2}\) (for which \(T_Z\) is a least-area enclosure), strong subadditivity (Lemma~\ref{lem:classical-submod}) implies that taking this union never increases the boundary area. We thus obtain a sweep \(T_p\to T_Z\) with peak area at most \(2\operatorname{Area}(T_p)\), which by \eqref{eq:two-sided-main-width} is strictly smaller than \(\operatorname{Area}(B[T_p,T_{\rm cut}])\). If \(T_Z=T_{\rm cut}\), this low-peak sweep from \(T_p\) to \(T_{\rm cut}\) already contradicts \eqref{eq:two-sided-main-width}. Otherwise, \(T_Z\in\mathcal E_{\rm out}[T_{\rm cut}]\) is a nontrivial exterior throat candidate for \(T_{\rm cut}\). Since concatenating the low-peak sweep \(T_p\to T_Z\) with any near-optimal sweep \(T_Z\to T_{\rm cut}\) produces a sweep \(T_p\to T_{\rm cut}\), the peak of the \(T_Z\to T_{\rm cut}\) sweep must be at least \(\operatorname{Area}(B[T_p,T_{\rm cut}])\):
\[
\operatorname{Area}\!\left(B[T_Z,T_{\rm cut}]\right)\ge \operatorname{Area}\!\left(B[T_p,T_{\rm cut}]\right).
\]
Combining this with the strict area inequality \eqref{eq:two-sided-z-small} yields
\[
\operatorname{Area}\!\left(B[T_Z,T_{\rm cut}]\right)-\operatorname{Area}(T_Z)
>\operatorname{Area}\!\left(B[T_p,T_{\rm cut}]\right)-\operatorname{Area}(T_p)
=\Delta[p],
\]
which contradicts the maximization property \eqref{eq:two-sided-old-max} of \(T_p\). Thus no enclosure of the two remnants inside \(D_{T_p}\) can have area below \(\operatorname{Area}(T_p)\). (Notice that \(T_Z=\partial(D_0\cup D_{T_{\rm cut}})\) is used here purely as an \emph{exterior throat candidate} in \(\mathcal E_{\rm out}[T_{\rm cut}]\), not as an inner cutoff in \(\mathcal A_p\), so it makes no difference if \(D_0\) and \(D_{T_{\rm cut}}\) touch \(p\) from opposite sides and their union contains \(p\) in its interior.)

\medskip\noindent\textbf{3. Exclude a common admissible cutoff and show that both sides are nonempty.}
Suppose a common admissible cutoff in \(\mathcal A_p\) enclosed \(D_{T_1}\cup D_{T_2}\), as in figure~\ref{fig:two-sided-cutoff}(a). Since both that cutoff and \(T_p\) are outer-minimal, a least-area enclosure \(D_0\) of \(D_{T_1}\cup D_{T_2}\) can be chosen inside both regions. Because \(D_{T_p}\) is itself an enclosing competitor, \(\operatorname{Area}(\partial D_0)\le\operatorname{Area}(T_p)\). However, since \(D_0\) lies inside an admissible cutoff in \(\mathcal A_p\), we have \(p\notin\operatorname{Int}D_0\), whereas Lemma~\ref{lem:cl-p-interior} gives \(p\in\operatorname{Int}D_{T_p}\). Thus \(\partial D_0\) and \(T_p\) are distinct throats, which forces the strict inequality \(\operatorname{Area}(\partial D_0)<\operatorname{Area}(T_p)\), contradicting Step~2. The same nondegeneracy argument shows that any minimizing enclosure of \(D_{T_1}\cup D_{T_2}\) inside \(D_{T_p}\) with boundary area equal to \(\operatorname{Area}(T_p)\) must coincide with \(D_{T_p}\) itself.

Consequently, neither \(T_1\) nor \(T_2\) can be empty: if \(D_{T_1}=\varnothing\), then \(D_{T_1}\cup D_{T_2}=D_{T_2}\) would already be enclosed by the single admissible cutoff \(T_2\in\mathcal A_p\) (and if both were empty, \(T_{\rm cut}=\varnothing\in\mathcal A_p\) would enclose them), a contradiction. Thus both \(T_1\) and \(T_2\) are nonempty, proving \(0<\operatorname{Area}(T_i)<\operatorname{Area}(T_p)\).

\medskip\noindent\textbf{4. Establish the strict sum bound.}
Finally, to establish the strict sum inequality \(\operatorname{Area}(T_1)+\operatorname{Area}(T_2)>\operatorname{Area}(T_p)\), apply Step~2 to the union \(D_0=D_{T_1}\cup D_{T_2}\subseteq D_{T_p}\) itself, which gives
\[
\operatorname{Area}\bigl(\partial(D_{T_1}\cup D_{T_2})\bigr)\ge\operatorname{Area}(T_p).
\]
If \(D_{T_1}\cup D_{T_2}\) is a proper subset of \(D_{T_p}\), the equality statement in Step~3 ensures that this inequality is strict, so \(\operatorname{Area}(T_1)+\operatorname{Area}(T_2)\ge\operatorname{Area}(\partial(D_{T_1}\cup D_{T_2}))>\operatorname{Area}(T_p)\). On the other hand, if \(D_{T_1}\cup D_{T_2}=D_{T_p}\), then the two side regions fill \(D_{T_p}\) and meet along a shared interface \(Q:=\partial D_{T_1}\cap\partial D_{T_2}\subseteq\gamma_{A_1}\) of strictly positive area passing through the interior point \(p\in\operatorname{Int}D_{T_p}\). Counting the shared interface on both sides then gives
\[
\operatorname{Area}(T_1)+\operatorname{Area}(T_2)
=\operatorname{Area}(T_p)+2\operatorname{Area}(Q)
>\operatorname{Area}(T_p).
\]
Thus the strict sum bound holds in all cases.
\end{proof}

\section{With quantum corrections}
\label{sec:quantum}

We now extend section~\ref{sec:classical} to the quantum-corrected regime by replacing \(\operatorname{Area}(\partial D)\) with \(4G_N S_{\rm gen}(D)\). 
To parallel section~\ref{sec:classical}, we first emphasize four differences introduced by bulk quantum fields:
\begin{enumerate}
\item \textbf{Nonvanishing entropy of the total system:}
As noted in section~\ref{sec:preliminaries}, evaluating the Python's Lunch complexity on the maximally mixed state \(\rho_{\rm code}\) of a nontrivial code subspace inevitably gives the boundary system \(A\) a nonzero entropy \(S(A)=\ln d_{\rm code}>0\). A key subtlety is that the exterior and interior generalized entropies \(S_{\rm gen}(O_X)\) and \(S_{\rm gen}(D_X)\) of a cut \(X\) no longer coincide exactly, so a surface that is minimal from one side need not be minimal from the other. To bound their discrepancy, we introduce a reference system \(P\) that purifies \(A\). Purity of the joint state on \(\Sigma \cup P\) implies \(S_{\rm bulk}(O_X)=S_{\rm bulk}(D_X\cup P)\), and since both sides share the same area term, we obtain
\begin{equation}\label{eq:code-entropy-sides}
S_{\rm gen}(O_X)-S_{\rm gen}(D_X)
  = S(A)-I_{\rm bulk}(D_X:P),\qquad
0\le I_{\rm bulk}(D_X:P)\le 2S(A),
\end{equation}
where \(I_{\rm bulk}(D_X:P)=S_{\rm bulk}(D_X)+S(P)-S_{\rm bulk}(D_X\cup P)\) is the bulk mutual information. Consequently, \(\bigl|S_{\rm gen}(O_X)-S_{\rm gen}(D_X)\bigr|\le S(A)=\ln d_{\rm code}\). Therefore, as long as all extremal surfaces being compared differ in generalized entropy by more than $2S(A)$, the offset $S(A)$ cannot affect any comparison. We further require that this offset not affect stationarity, so that $A_1$ and $A_{\bar{1}}$ share (approximately) the same extremal surfaces. We therefore restrict the code subspace to be small enough that it affects neither comparisons nor extremality, and we treat $S(A)$ as negligible throughout this section. Corrections at finite \(S(A)\) are discussed in section~\ref{sec:discussion}.

\item \textbf{Bulk entanglement entropy versus boundary area:} Unlike geometric area, the bulk matter von Neumann entropy \(S_{\rm bulk}(D)\) is nonlocal and nonadditive. Strictly speaking, defining a subsystem entanglement entropy in full quantum gravity is subtle due to diffeomorphism invariance and gravitational nonlocality; here we work in the standard regulated semiclassical approximation in which \(S_{\rm bulk}(D)\) is well-defined for bulk spatial regions \(D\subseteq\Sigma\). Fortunately, the bulk entanglement entropy satisfies strong subadditivity, analogous to Lemma~\ref{lem:classical-submod} for the boundary area:
\begin{equation}
S_{\rm bulk}(D\cap D')+S_{\rm bulk}(D\cup D')
\le S_{\rm bulk}(D)+S_{\rm bulk}(D').
\end{equation}
Combining both properties, we obtain the strong subadditivity of generalized entropy:
\begin{lemma}[Strong Subadditivity of Generalized Entropy]
\label{lem:quantum-submod}
For any two bulk regions \(D,D'\subseteq\Sigma\) evaluated in the same state with their prescribed reference systems,
\begin{equation}\label{eq:q-pure-submod}
S_{\rm gen}(D\cap D')+S_{\rm gen}(D\cup D')
\le S_{\rm gen}(D)+S_{\rm gen}(D').
\end{equation}
\end{lemma}
\item \textbf{Matter-entropy variation during contractions:} Under a small inward deformation,\footnote{We measure deformation size by \(d_{\rm vol}(D,D'):=\operatorname{Vol}(D\triangle D')\), where \(D\triangle D'=(D\setminus D')\cup(D'\setminus D)\).}\phantomsection\label{def:inf-deform} \(S_{\rm bulk}(D)\) can increase even when \(\operatorname{Area}(\partial D)\) decreases. To prevent uncontrolled spikes in generalized entropy during a contraction, we impose the following requirement on \(S_{\rm bulk}\).
\begin{assumption}[Volumetric Lipschitz Bound on Bulk Matter Entropy]
\label{assump:quantum-lipschitz}
For all bulk regions \(D,D'\subseteq\Sigma\) used in our constructions, there exists a finite constant \(K_{\rm ent}>0\) such that under any deformation,
\begin{equation}\label{eq:q-pure-entropy-bound}
\bigl|S_{\rm bulk}(D)-S_{\rm bulk}(D')\bigr|
\le K_{\rm ent}\,d_{\rm vol}(D,D')
=K_{\rm ent}\operatorname{Vol}(D\triangle D').
\end{equation}
Physically, \eqref{eq:q-pure-entropy-bound} expresses the requirement that in a regulated semiclassical state, adding or removing a small bulk spatial volume \(\Delta\operatorname{Vol}\) changes the bulk matter entropy by at most \(O(\Delta\operatorname{Vol})\), without requiring \(S_{\rm bulk}\) to be extensive or additive.
\end{assumption}
\item \textbf{Entanglement islands and nonlocal stability:}\phantomsection\label{rem:islands-vs-closed} While closed RT components not reaching the asymptotic boundary already exist classically (see section~\ref{sec:classical}), the quantum regime introduces \emph{entanglement islands} --- disconnected bulk regions in \(\operatorname{EW}(A_1)\) bounded by closed quantum extremal surfaces (QESs) that are stabilized by bulk entanglement (e.g., with a nongravitating bath). Moreover, cutting a quantum throat along a closed QES or island boundary changes the global region and its bulk entanglement partners, so the resulting piece is not automatically a quantum throat even when no geometric corners are created.
\end{enumerate}

In addition, we use the quantum entanglement shadow \(\mathcal S_{\rm ent}^{\rm Q}(A)\) in \eqref{eq:prelim-shadow}, the quantum admissible cutoff class \(\mathcal A_p^{\rm Q}\), and the quantum point-decoding complexity \(\Delta[p]\) in \eqref{eq:pl-review-point-quantum} of Definition~\ref{def:classical-point-complexity}, together with the generic nondegeneracy of Assumption~\ref{assump:no-coincidence}. Precisely, we define a \emph{quantum throat}\phantomsection\label{def:quantum-throat} as the boundary \(T=\partial D_T\) of a compact retained region \(D_T\) that is a local minimum of \(S_{\rm gen}\) under small volumetric deformations,\footnote{There is a neighborhood \(d_{\rm vol}(D',D_T)\le\varepsilon_{\rm vol}\) in which every admitted \(D'\) obeys \(S_{\rm gen}(D')\ge S_{\rm gen}(D_T)\).} and call \(T\) \emph{outer-minimal} if \(S_{\rm gen}(D')\ge S_{\rm gen}[T]\) for every admitted enclosure \(D'\supseteq D_T\). For \(\Delta[p]>0\), the exact same argument as in Lemma~\ref{lem:cl-p-interior} (using Lemma~\ref{lem:quantum-submod} in place of Lemma~\ref{lem:classical-submod}) shows that the maximizing outer quantum throat \(T_p\) is outer-minimal and satisfies \(p\in\operatorname{Int}D_{T_p}\), while Assumption~\ref{assump:no-coincidence} ensures that all optimal cutoffs select the same dominant pair \((B_p,T_p)\).

\subsection{Discrete variational contraction and fixed-throat shadow theorem}
\label{sec:quantum-fixed}

Under quantum corrections, an arbitrary geometric contraction need not monotonically decrease \(S_{\rm gen}\), which requires a careful revision of the original protocol. We first construct a discrete variational contraction and then interpolate between successive regions.

\paragraph{Discrete variational contraction.}
\phantomsection\label{par:q-contraction}
Let \(D^{(0)}\) be a compact locally outward-minimizing region. For a fixed \(\varepsilon_{\rm vol}>0\), choose successive inward contractions by\footnote{We assume existence of the variational minima and least-entropy enclosures in the admitted class. At fixed regulator, compact comparison domains, a closed finite-perimeter class, and continuity of the matter entropy provide a sufficient setting, with regular terminal boundaries assumed.}
\begin{equation}\label{eq:q-pure-deletion}
D^{(k+1)}\in\operatorname*{arg\,min}_{\substack{
 D'\subseteq D^{(k)}\\
 d_{\rm vol}(D^{(k)},D')\le\varepsilon_{\rm vol}}}
 S_{\rm gen}(D')
\qquad(k=0,1,\ldots).
\end{equation}
This produces a nested sequence with nonincreasing generalized entropy and preserves outward minimality. In particular,
\begin{equation}\label{eq:q-pure-step-outer}
S_{\rm gen}(D^{(k)})\le S_{\rm gen}(D')\qquad(D^{(k)}\subseteq D'\subseteq D^{(0)}).
\end{equation}
If \(D^{(0)}\) is globally outward-minimizing, that stronger property is preserved as well.

To verify \eqref{eq:q-pure-step-outer} by induction on \(k\), intersect any proposed enclosure \(D'\) (\(D^{(k+1)}\subseteq D'\subseteq D^{(0)}\)) with \(D^{(k)}\). Then \(D'\cap D^{(k)}\subseteq D^{(k)}\) satisfies \(d_{\rm vol}(D^{(k)},D'\cap D^{(k)})\le d_{\rm vol}(D^{(k)},D^{(k+1)})\le\varepsilon_{\rm vol}\), so \(D'\cap D^{(k)}\) is an allowed competitor in \eqref{eq:q-pure-deletion}, giving \(S_{\rm gen}(D^{(k+1)})\le S_{\rm gen}(D'\cap D^{(k)})\). Combining this with the inductive outward comparison \(S_{\rm gen}(D^{(k)})\le S_{\rm gen}(D'\cup D^{(k)})\) and Lemma~\ref{lem:quantum-submod} yields \(S_{\rm gen}(D^{(k+1)})\le S_{\rm gen}(D')\), inheriting the initial local (or global) outward minimality of \(D^{(0)}\).

The iteration stops when a minimizing step removes strictly less volume than \(\varepsilon_{\rm vol}\) (possibly at \(D^{(k)}=\varnothing\)). If the terminal region \(D_{{\rm end},\varepsilon_{\rm vol}}\neq\varnothing\), the unused volume allowance implies that no inward \(\varepsilon'\)-contraction (for sufficiently small \(\varepsilon'>0\)) can lower \(S_{\rm gen}\). Combined with the inherited local or global outward minimality and Lemma~\ref{lem:quantum-submod}, this shows that \(D_{{\rm end},\varepsilon_{\rm vol}}\) is a genuine local minimum of \(S_{\rm gen}\), so \(\partial D_{{\rm end},\varepsilon_{\rm vol}}\) is a quantum throat.

\phantomsection\label{par:q-zero-step}
Next, we connect the discrete steps \(D^{(k)}\supseteq D^{(k+1)}\) into a continuous nested sweep. The nested area interpolation lemma~\cite[Lemma~1.12]{ChodoshLiokumovichSpolaor2022} applies to finite-perimeter endpoints with uniformly bounded boundary area. Together with Assumption~\ref{assump:quantum-lipschitz}, it ensures that for any prescribed tolerance \(\varepsilon_{\rm ent}>0\) and sufficiently small step volume \(\varepsilon_{\rm vol}>0\), each discrete pair \((D^{(k)},D^{(k+1)})\) can be joined by an admissible continuous nested sweep \(\Omega_k(s)\) (\(\Omega_k(0)=D^{(k)}\), \(\Omega_k(1)=D^{(k+1)}\)) satisfying
\begin{equation}\label{eq:q-interpolation-peak}
\max_{s\in[0,1]}S_{\rm gen}\bigl(\Omega_k(s)\bigr)
\le S_{\rm gen}\bigl(D^{(k)}\bigr)+\delta,
\qquad
\delta:=\varepsilon_{\rm ent}+2K_{\rm ent}\varepsilon_{\rm vol}.
\end{equation}

In all subsequent comparisons, we take \(\delta\to 0\) and omit these infinitesimal excess terms. Furthermore, when we take the union of the resulting sweeps for two disjoint starting pieces \(D_1,D_2\), strong subadditivity (Lemma~\ref{lem:quantum-submod}) and the monotone discrete contractions yield a combined sweep peak bounded by \(S_{\rm gen}(D_1)+S_{\rm gen}(D_2)\).

Unlike the area functional, whose first variation is local, generalized entropy includes the nonlocal contribution $S_{\rm bulk}$. One might wonder whether a quantum throat that halts one monotone contraction could be bypassed by choosing a different contraction path. We therefore establish two lemmas showing that quantum throats can be retained by a suitable choice of variational contraction.

\begin{lemma}[Variational Contractions Retain Interior Quantum Throats]
\label{lem:q-avoidance}
Suppose a fixed quantum throat region \(D_{T'}\) lies inside a contracting region \(D^{(k)}\) (\(D_{T'}\subseteq D^{(k)}\)). For \(\varepsilon_{\rm vol}>0\) chosen within the local-minimum neighborhood of \(D_{T'}\), every deletion step \eqref{eq:q-pure-deletion} can be chosen to retain \(D_{T'}\) (\(D_{T'}\subseteq D^{(k+1)}\)). In particular, we may first contract a smaller region to a stable remnant and subsequently contract an enclosing region while preserving that remnant inside the new terminal region.
\end{lemma}

\begin{proof}
Take \(\varepsilon_{\rm vol}>0\) small enough that \(S_{\rm gen}(D'')\ge S_{\rm gen}(D_{T'})\) for every allowed region \(D''\) satisfying \(d_{\rm vol}(D'',D_{T'})=\operatorname{Vol}(D''\triangle D_{T'})\le\varepsilon_{\rm vol}\). Let \(D'\subseteq D^{(k)}\) be any candidate inward \(\varepsilon_{\rm vol}\)-contraction of \(D^{(k)}\), so \(\operatorname{Vol}(D^{(k)}\setminus D')\le\varepsilon_{\rm vol}\). Although \(D'\) itself need not be close to \(D_{T'}\) in volume, its intersection \(D'\cap D_{T'}\) differs from \(D_{T'}\) only inside \(D^{(k)}\setminus D'\):
\[
D_{T'}\setminus D'\subseteq D^{(k)}\setminus D'
\quad\Longrightarrow\quad
d_{\rm vol}(D'\cap D_{T'},D_{T'})
=\operatorname{Vol}\bigl((D'\cap D_{T'})\triangle D_{T'}\bigr)
\le\varepsilon_{\rm vol}.
\]
Local minimality of \(D_{T'}\) therefore gives \(S_{\rm gen}(D'\cap D_{T'})\ge S_{\rm gen}(D_{T'})\). Applying Lemma~\ref{lem:quantum-submod} to \(D'\) and \(D_{T'}\) then yields
\[
S_{\rm gen}(D'\cup D_{T'})
\le S_{\rm gen}(D')+S_{\rm gen}(D_{T'})-S_{\rm gen}(D'\cap D_{T'})
\le S_{\rm gen}(D').
\]
Since \(D_{T'}\subseteq D'\cup D_{T'}\subseteq D^{(k)}\) and \(\operatorname{Vol}(D^{(k)}\setminus(D'\cup D_{T'}))\le\operatorname{Vol}(D^{(k)}\setminus D')\le\varepsilon_{\rm vol}\), the restored region \(D'\cup D_{T'}\) is an admissible \(\varepsilon_{\rm vol}\)-contraction step with generalized entropy no larger than \(S_{\rm gen}(D')\).
\end{proof}

\begin{lemma}[Induced Contraction from Intermediate Regions]
\label{lem:q-halfway}
Suppose a chosen discrete contraction \(\{D^{(k)}\}\) takes \(D^{(0)}\) to a terminal region \(D_{\rm end}\). For every intermediate starting region \(D'\) satisfying \(D_{\rm end}\subseteq D'\subseteq D^{(0)}\), the nested sequence \(D^{(k)}\cap D'\) reaches the same terminal region with nonincreasing generalized entropy. Interpolation gives the corresponding admissible sweep with the peak control described above.
\end{lemma}

\begin{proof}
The intersection sequence starts at \(D'\), ends at \(D_{\rm end}\), and is nested. Lemma~\ref{lem:quantum-submod} gives
\[
\begin{aligned}
S_{\rm gen}\bigl(D^{(k+1)}\cap D'\bigr)
&\le S_{\rm gen}\bigl(D^{(k)}\cap D'\bigr)+S_{\rm gen}\bigl(D^{(k+1)}\bigr)\\
&\quad-S_{\rm gen}\Bigl(D^{(k+1)}\cup\bigl(D^{(k)}\cap D'\bigr)\Bigr)\\
&\le S_{\rm gen}\bigl(D^{(k)}\cap D'\bigr).
\end{aligned}
\]
The second inequality uses \eqref{eq:q-pure-step-outer}, since the union encloses \(D^{(k+1)}\) inside \(D^{(0)}\). Moreover, each intersection step removes volume at most \(\operatorname{Vol}(D^{(k)}\setminus D^{(k+1)})\le\varepsilon_{\rm vol}\), while the boundary areas \(\operatorname{Area}\bigl(\partial(D^{(k)}\cap D')\bigr)\) remain bounded independently of \(k\) (since \(S_{\rm gen}(D^{(k)}\cap D')\le S_{\rm gen}(D')\) and \(S_{\rm bulk}\) is controlled by Assumption~\ref{assump:quantum-lipschitz}). Applying the area interpolation lemma~\cite[Lemma~1.12]{ChodoshLiokumovichSpolaor2022} and Assumption~\ref{assump:quantum-lipschitz} to the successive intersections, then taking \(\delta\to 0\) as above, yields an admissible continuous sweep from \(D'\) to \(D_{\rm end}\) with peak at most \(S_{\rm gen}(D')\).
\end{proof}

This construction yields the quantum counterpart of Theorem~\ref{thm:cl-fixed}.
\begin{theorem}[Quantum Fixed-Throat Entanglement Shadow Theorem]
\label{thm:q-fixed}
Let \(T\) be a quantum throat in the pure-state bulk slice \(\Sigma\). Under Assumptions~\ref{assump:no-coincidence} and~\ref{assump:quantum-lipschitz} and the variational setting above, suppose no nonempty quantum throat inside \(D_{T}\) has generalized entropy strictly smaller than \(S_{\rm gen}[T]\). Then the minimax bulge \(B[T,\varnothing]\) inside \(T\) satisfies
\begin{equation}\label{eq:q-pure-shadow-threshold}
\boxed{\quad
S_{\rm gen}\!\bigl[B[T,\varnothing]\bigr]>2S_{\rm gen}[T]
\quad\Longrightarrow\quad
\operatorname{Int}D_{T}\subseteq\mathcal S_{\rm ent}^{\rm Q}(A).
\quad}
\end{equation}
\end{theorem}
\begin{proof}
Suppose a globally minimizing probe QES \(\gamma_{A_1}\), including any closed QES or island boundary, enters \(\operatorname{Int}D_T\).

\medskip\noindent\textbf{1. Bound the two side regions.}
Define the two pieces (as in figure~\ref{fig:cutting-sweep}) by
\begin{equation}\label{eq:q-cut-pieces}
D_1:=D_T\cap\operatorname{EW}(A_1),\qquad
D_2:=D_T\cap\operatorname{EW}(A_1)^c.
\end{equation}
To bound \(S_{\rm gen}(D_1)\), adjoin \(D_T\) to \(\operatorname{EW}(A_1)\) to form the candidate entanglement wedge \(D_T\cup\operatorname{EW}(A_1)\). Global minimality of \(\operatorname{EW}(A_1)\) gives \(S_{\rm gen}(D_T\cup\operatorname{EW}(A_1))\ge S_{\rm gen}(\operatorname{EW}(A_1))\); applying Lemma~\ref{lem:quantum-submod} to \(D=D_T\) and \(D'=\operatorname{EW}(A_1)\) then yields \(S_{\rm gen}(D_1)\le S_{\rm gen}(D_T)=S_{\rm gen}[T]\). Purity of the total state gives the exact same comparison for \(\operatorname{EW}(A_1)^c\), proving \(S_{\rm gen}(D_2)\le S_{\rm gen}[T]\) without ever assigning a standalone entropy to the surface fragment \(\gamma_{A_1}\cap D_T\).

Thus
\begin{equation}\label{eq:q-pure-side-budget}
S_{\rm gen}(D_1)\le S_{\rm gen}[T],\qquad
S_{\rm gen}(D_2)\le S_{\rm gen}[T].
\end{equation}

\medskip\noindent\textbf{2. Check outward minimality and contract.}
Let \(D'\supseteq D_1\) be an enclosing region sufficiently close to \(D_1\) in volume. Then the union \(D_T\cup D'\) is close in volume to \(D_T\) because
\[
\operatorname{Vol}\bigl((D_T\cup D')\triangle D_T\bigr)
=\operatorname{Vol}(D'\setminus D_T)
\le\operatorname{Vol}(D'\setminus D_1).
\]
Local minimality of the quantum throat \(T\) therefore gives \(S_{\rm gen}(D_T\cup D')\ge S_{\rm gen}(D_T)\). Applying Lemma~\ref{lem:quantum-submod} first to \(D_T\) and \(D'\) gives \(S_{\rm gen}(D_T\cap D')\le S_{\rm gen}(D')\), and applying Lemma~\ref{lem:quantum-submod} next to \(D_T\cap D'\) and the globally minimizing wedge \(\operatorname{EW}(A_1)\) gives \(S_{\rm gen}(D_1)\le S_{\rm gen}(D_T\cap D')\). The proof for \(D_2\) is identical, so for either side and every sufficiently close enclosing region \(D'\supseteq D_i\),
\begin{equation}\label{eq:q-piece-initial-outward}
S_{\rm gen}(D_i)\le S_{\rm gen}(D_T\cap D')\le S_{\rm gen}(D').
\end{equation}
When \(T\) is outer-minimal, \(S_{\rm gen}(D_T\cup D')\ge S_{\rm gen}(D_T)\) holds for every enclosing region \(D'\) without any closeness restriction, so \eqref{eq:q-piece-initial-outward} holds globally. Because each piece \(D_i\) is locally outward-minimizing, we may apply the discrete variational contraction above: each side therefore either terminates at a quantum throat \(T_i\) or contracts to the empty set.

\medskip\noindent\textbf{3. Exclude nonempty remnants and compare with the threshold.}
Each terminal throat satisfies \(S_{\rm gen}[T_i]\le S_{\rm gen}(D_i)\le S_{\rm gen}[T]\). Because each nonempty \(D_{T_i}\subseteq D_i\subsetneq D_T\) is distinct from \(D_T\), generic nondegeneracy forces \(S_{\rm gen}[T_i]<S_{\rm gen}[T]\), contrary to the hypothesis that there is no smaller quantum throat inside \(D_T\). Thus both stopping regions are empty, and the two-side sweep bound gives
\begin{equation}\label{eq:q-pure-cut-peak}
S_{\rm gen}\!\bigl[B[T,\varnothing]\bigr]
\le S_{\rm gen}(D_1)+S_{\rm gen}(D_2)
\le 2S_{\rm gen}[T],
\end{equation}
contradicting \eqref{eq:q-pure-shadow-threshold}. Exclusion of all probe QESs from the open interior \(\operatorname{Int}D_T\) proves \(\operatorname{Int}D_T\subseteq\mathcal S_{\rm ent}^{\rm Q}(A)\).
\end{proof}

\begin{remark}
\label{rem:q-threshold-shift}
Comparing Theorem~\ref{thm:q-fixed} with the classical Theorem~\ref{thm:cl-fixed}, the threshold for the generalized entropy of the interior bulge \(B[T,\varnothing]\) is shifted by the bulk matter entropy of the retained region:
\begin{equation}\label{eq:q-pure-area-correction}
2\times 4G_NS_{\rm gen}[T]
=2\operatorname{Area}(T)+8G_NS_{\rm bulk}(D_{T}).
\end{equation}
Thus a bulge area \(\operatorname{Area}(B[T,\varnothing])\) exceeding \(2\operatorname{Area}(T)\) alone is no longer sufficient to guarantee that \(\operatorname{Int}D_{T}\) lies in the quantum entanglement shadow; the generalized entropy \(S_{\rm gen}\!\bigl[B[T,\varnothing]\bigr]\) of the interior bulge must exceed \(2S_{\rm gen}[T]\).
\end{remark}

\subsection{The quantum point-decoding theorem}
\label{sec:quantum-comparison-lemmas}

To prove the quantum version of the point-decoding theorem, we first establish the remaining quantum enclosure and cutoff-replacement lemmas paralleling section~\ref{sec:classical}. Because \(S_{\rm bulk}\) is nonlocal, we establish these properties along the discrete variational contraction using the strong subadditivity of generalized entropy (Lemma~\ref{lem:quantum-submod}).

\begin{lemma}[Least-Entropy Enclosure of Stable Remnants]
\label{lem:q-enclosure}
Let \(D_1,D_2\) be two quantum throat regions, and let \(D_E\) be a least-entropy region enclosing \(D_1\cup D_2\). Then \(\partial D_E\) is an outer-minimal quantum throat, and \(D_E\) can be chosen inside any outer-minimal region that already encloses \(D_1\cup D_2\).
\end{lemma}

\begin{proof}
First, the union \(D_1\cup D_2\) is locally inward-minimizing: for any \(D'\subseteq D_1\cup D_2\) with \(d_{\rm vol}(D',D_1\cup D_2)\le\varepsilon_{\rm vol}\), applying Lemma~\ref{lem:quantum-submod} and local minimality successively to \(D_1\) and \(D_2\) gives
\[
S_{\rm gen}(D')\ge S_{\rm gen}(D'\cup D_1)\ge S_{\rm gen}(D_1\cup D_2).
\]
Next, let \(D''\) be any region within an \(\varepsilon_{\rm vol}\)-deformation of \(D_E\). Because \(D''\cap(D_1\cup D_2)\) is within \(\varepsilon_{\rm vol}\) of \(D_1\cup D_2\), the local minimality just shown and Lemma~\ref{lem:quantum-submod} imply
\[
S_{\rm gen}(D'')\ge S_{\rm gen}(D''\cup D_1\cup D_2)\ge S_{\rm gen}(D_E),
\]
where the second inequality holds because \(D''\cup D_1\cup D_2\) is an enclosure of \(D_1\cup D_2\) and \(D_E\) minimizes \(S_{\rm gen}\) over all such enclosures. Thus \(\partial D_E\) is locally minimizing (even under variations that pass through \(D_1\cup D_2\)) and outer-minimal. Finally, intersecting \(D_E\) with any outer-minimal region enclosing \(D_1\cup D_2\) does not increase \(S_{\rm gen}\) by Lemma~\ref{lem:quantum-submod}.
\end{proof}

In particular, applying Lemma~\ref{lem:q-enclosure} to a single quantum throat \(T\) shows that outer-minimality against all enclosing regions \(D'\supseteq D_T\) is equivalent to requiring \(S_{\rm gen}[T']\ge S_{\rm gen}[T]\) for every enclosing quantum throat \(T'\). Next, we prove the quantum version of Lemma~\ref{lem:cl-cutoff-replace}.

\begin{lemma}[Least-Entropy Enclosure as the Optimal Cutoff]
\label{lem:q-cutoff-replace}
Suppose \(\Delta[p]>4G_NS_{\rm gen}[T_p]\) and \(p\in\gamma_{A_1}\). Let \(T_{\rm cut}\in\mathcal A_p^{\rm Q}\) be an optimal cutoff for \(\Delta[p]\). Cut \(D_{T_{\rm cut}}\) along \(\gamma_{A_1}\), contract both sides via the variational contraction above, and let \(\widehat T_{\rm cut}=\partial D_{\widehat T_{\rm cut}}\) with \(D_{\widehat T_{\rm cut}}\subseteq D_{T_{\rm cut}}\) be a least-entropy enclosure of their terminal remnants \(D_R\). Then \(\widehat T_{\rm cut}\in\mathcal A_p^{\rm Q}\) is also an optimal cutoff for \(\Delta[p]\) with the same dominant pair \((B_p,T_p)\).
\end{lemma}

\begin{proof}
By Lemma~\ref{lem:q-enclosure}, \(\widehat T_{\rm cut}\) is an outer-minimal quantum throat that can be chosen inside \(D_{T_{\rm cut}}\), so \(p\notin\operatorname{Int}D_{\widehat T_{\rm cut}}\) and \(\widehat T_{\rm cut}\in\mathcal A_p^{\rm Q}\). Just as in Lemma~\ref{lem:cl-cutoff-replace}, it suffices to consider an arbitrary outer-minimal exterior candidate \(T_X\in\mathcal E_{\rm out}^{\rm Q}[\widehat T_{\rm cut}]\). Following the two-part sweep construction of Lemma~\ref{lem:cl-cutoff-replace}, Lemma~\ref{lem:quantum-submod} and Lemma~\ref{lem:q-halfway} guarantee in Part~2 that each side of \(D_{T_X}\cap D_{T_{\rm cut}}\) contracts to the same side remnant in \(D_R\) as \(D_{T_{\rm cut}}\) did, with the same minimax peak bound. Combining the two parts yields
\[
S_{\rm gen}\!\bigl[B[T_X,\widehat T_{\rm cut}]\bigr]-S_{\rm gen}[T_X]
\le\max\left\{\frac{\Delta[p]}{4G_N},\,S_{\rm gen}[T_{\rm cut}]\right\}
=\frac{\Delta[p]}{4G_N},
\]
where the equality uses \(S_{\rm gen}[T_{\rm cut}]\le S_{\rm gen}[T_p]<\Delta[p]/(4G_N)\) from outer-minimality of \(T_{\rm cut}\). Hence \(\widehat T_{\rm cut}\) is also optimal; Assumption~\ref{assump:no-coincidence} identifies its dominant pair with \((B_p,T_p)\).
\end{proof}

\begin{theorem}[Quantum Point-Decoding Entanglement Shadow Theorem]
\label{thm:q-point}
Under Assumptions~\ref{assump:no-coincidence} and~\ref{assump:quantum-lipschitz}, suppose a bulk point \(p\in\Sigma\) satisfies
\begin{equation}\label{eq:q-pure-point-threshold}
\Delta[p]>4G_NS_{\rm gen}[T_p].
\end{equation}
If \(p\in\gamma_{A_1}\) for a globally minimizing probe QES \(\gamma_{A_1}\) (including any closed QES or entanglement island boundary), then there exist two nonempty quantum throats \(T_1,T_2\), with
\begin{equation}\label{eq:q-remnant-locations}
D_{T_1}\subseteq D_{T_p}\cap\operatorname{EW}(A_1),\qquad
D_{T_2}\subseteq D_{T_p}\cap\operatorname{EW}(A_1)^c,
\end{equation}
such that:
\begin{enumerate}
\item Each throat \(T_i\) (\(i=1,2\)) is an admissible quantum cutoff in \(\mathcal A_p^{\rm Q}\), but no admissible cutoff in \(\mathcal A_p^{\rm Q}\) encloses \(D_{T_1}\cup D_{T_2}\).
\item Their individual generalized entropies and the generalized entropy of their union satisfy
\begin{equation}\label{eq:q-pure-two-side-sizes}
0<S_{\rm gen}[T_i]<S_{\rm gen}[T_p]\qquad(i=1,2),
\end{equation}
and
\begin{equation}\label{eq:q-pure-union-claim}
\boxed{\quad
S_{\rm gen}(D_{T_1}\cup D_{T_2})\ge S_{\rm gen}[T_p],
\quad}
\end{equation}
where \eqref{eq:q-pure-union-claim} is strict whenever \(D_{T_1}\cup D_{T_2}\ne D_{T_p}\).
\end{enumerate}
\end{theorem}

\begin{proof} The proof parallels that of Theorem~\ref{thm:cl-point}.

\medskip\noindent\textbf{1. Construct the two-sided stopping throats.}
By Lemma~\ref{lem:q-cutoff-replace}, we choose the least-entropy enclosure of the cutoff remnants \(D_R\) as the optimal cutoff, and denote it by \(T_{\rm cut}\). Cut \(D_{T_p}\) along \(\gamma_{A_1}\) and contract both sides via the variational contraction above, while invoking Lemma~\ref{lem:q-avoidance} to retain \(D_R\subseteq D_{T_{\rm cut}}\) inside the resulting terminal regions, so that \(D_R\subseteq D_{T_1}\cup D_{T_2}\). Either stopping throat may initially be empty. Each nonempty \(T_i\) is outer-minimal and satisfies \(S_{\rm gen}[T_i]<S_{\rm gen}[T_p]\), and the sweep from \(T_p\) to \(\partial(D_{T_1}\cup D_{T_2})\) has peak generalized entropy at most \(2S_{\rm gen}[T_p]\).

The two remnants lie on opposite sides of \(\gamma_{A_1}\), as in \eqref{eq:q-remnant-locations}. Since \(p\in\gamma_{A_1}\), each excludes \(p\) from its open interior, although its boundary may touch \(p\). Thus each nonempty \(T_i\) belongs to \(\mathcal A_p^{\rm Q}\).

\medskip\noindent\textbf{2. Rule out a low-entropy enclosure of both remnants.}
Suppose an enclosure of \(D_{T_1}\cup D_{T_2}\) inside \(D_{T_p}\) has generalized entropy strictly below \(S_{\rm gen}[T_p]\). By Lemma~\ref{lem:q-enclosure}, choose a least-entropy enclosure \(D_0\subseteq D_{T_p}\); its boundary is an outer-minimal quantum throat, and \(S_{\rm gen}(D_0)<S_{\rm gen}[T_p]\).

Because \(D_0\cap D_{T_{\rm cut}}\) still encloses \(D_R\), the least-entropy property of \(T_{\rm cut}\) gives \(S_{\rm gen}(D_0\cap D_{T_{\rm cut}})\ge S_{\rm gen}[T_{\rm cut}]\). Strong subadditivity (Lemma~\ref{lem:quantum-submod}) therefore yields
\[
S_{\rm gen}(D_0\cup D_{T_{\rm cut}})
\le S_{\rm gen}(D_0)<S_{\rm gen}[T_p].
\]
The union remains inside \(D_{T_p}\) and encloses both remnants, so it is also a least-entropy enclosure. Hence Lemma~\ref{lem:q-enclosure} makes \(T_Z:=\partial(D_0\cup D_{T_{\rm cut}})\) a quantum throat enclosing \(T_{\rm cut}\), with \(S_{\rm gen}[T_Z]<S_{\rm gen}[T_p]\).

As in Step~2 of the classical proof, take the union of every stage of the Step~1 sweep with \(D_{T_Z}\). Its intersection with \(D_{T_Z}\) still encloses both remnants; the least-entropy property of \(D_{T_Z}\) and strong subadditivity show that this operation never increases generalized entropy. The resulting \(T_p\to T_Z\) sweep has peak at most \(2S_{\rm gen}[T_p]<S_{\rm gen}\!\bigl[B[T_p,T_{\rm cut}]\bigr]\). If \(T_Z=T_{\rm cut}\), this already contradicts the original minimax height. Otherwise, concatenation with a near-optimal \(T_Z\to T_{\rm cut}\) sweep forces
\[
S_{\rm gen}\!\bigl[B[T_Z,T_{\rm cut}]\bigr]\ge S_{\rm gen}\!\bigl[B[T_p,T_{\rm cut}]\bigr].
\]
Together with the strictly smaller throat entropy, this gives
\[
S_{\rm gen}\!\bigl[B[T_Z,T_{\rm cut}]\bigr]-S_{\rm gen}[T_Z]
>S_{\rm gen}\!\bigl[B[T_p,T_{\rm cut}]\bigr]-S_{\rm gen}[T_p]
=\frac{\Delta[p]}{4G_N},
\]
contradicting the exterior maximization for \(T_{\rm cut}\). Thus no such low-entropy enclosure exists. Here \(T_Z\) is only an exterior throat candidate; it need not exclude \(p\) from its interior.

\medskip\noindent\textbf{3. Exclude a common admissible cutoff and show that both sides are nonempty.}
Suppose a common cutoff in \(\mathcal A_p^{\rm Q}\) enclosed \(D_{T_1}\cup D_{T_2}\). By Lemma~\ref{lem:q-enclosure}, a least-entropy enclosure \(D_0\) can be chosen inside both that cutoff and \(D_{T_p}\). Since \(D_{T_p}\) is an enclosing competitor, \(S_{\rm gen}(D_0)\le S_{\rm gen}[T_p]\). However, \(p\notin\operatorname{Int}D_0\), whereas \(p\in\operatorname{Int}D_{T_p}\). These are distinct throats, so the assumed nondegeneracy gives \(S_{\rm gen}(D_0)<S_{\rm gen}[T_p]\), contradicting Step~2. The same nondegeneracy argument shows that any minimizing enclosure inside \(D_{T_p}\) with entropy equal to \(S_{\rm gen}[T_p]\) must coincide with \(D_{T_p}\).

Neither side can therefore be empty: otherwise the other admissible throat would itself enclose both remnants; if both were empty, the empty cutoff would do so. Thus \(T_1,T_2\) are nonempty, and Step~1 supplies their strict upper entropy bounds.

\medskip\noindent\textbf{4. Establish the union-entropy bound and the individual lower bounds.}
Apply Step~2 of this proof to \(D_{T_1}\cup D_{T_2}\) itself to obtain
\[
S_{\rm gen}(D_{T_1}\cup D_{T_2})\ge S_{\rm gen}[T_p].
\]
If equality holds, the union is itself a least-entropy enclosure of both remnants. Lemma~\ref{lem:q-enclosure} makes its boundary a quantum throat, and the equality statement in Step~3 forces \(D_{T_1}\cup D_{T_2}=D_{T_p}\). This proves \eqref{eq:q-pure-union-claim}, including strictness for a proper subset.

Finally, strong subadditivity gives
\[
S_{\rm gen}[T_1]+S_{\rm gen}[T_2]
\ge S_{\rm gen}(D_{T_1}\cup D_{T_2})
\ge S_{\rm gen}[T_p].
\]
Since each individual entropy is strictly below \(S_{\rm gen}[T_p]\), the other must be strictly positive. Hence \(0<S_{\rm gen}[T_i]<S_{\rm gen}[T_p]\) for \(i=1,2\), completing \eqref{eq:q-pure-two-side-sizes}. Unlike the classical area argument, the joint bound is stated for the entropy of the union, without assuming additivity of generalized entropy.
\end{proof}

\begin{remark}
\label{rem:quantum-union-vs-sum}
In the classical Theorem~\ref{thm:cl-point}, the joint constraint \eqref{eq:cl-two-sided-area-bounds} was written directly in terms of the sum of the two throat areas, \(\operatorname{Area}(T_1)+\operatorname{Area}(T_2)>\operatorname{Area}(T_p)\). In the quantum Theorem~\ref{thm:q-point}, the quantity naturally lower-bounded by \(S_{\rm gen}[T_p]\) is the generalized entropy of the union \(S_{\rm gen}(D_{T_1}\cup D_{T_2})\). To relate \(S_{\rm gen}(D_{T_1}\cup D_{T_2})\) to the individual generalized entropies \(S_{\rm gen}[T_1]\) and \(S_{\rm gen}[T_2]\), note that they can share an internal interface \(\partial D_{T_1}\cap\partial D_{T_2}\subseteq\gamma_{A_1}\). Because \(\operatorname{Int}D_{T_1}\) and \(\operatorname{Int}D_{T_2}\) are disjoint, we have the exact decomposition
\begin{equation}\label{eq:q-pure-mutual-interface}
S_{\rm gen}(D_{T_1}\cup D_{T_2})
=S_{\rm gen}[T_1]+S_{\rm gen}[T_2]
-I_{\rm bulk}(D_{T_1}:D_{T_2})
-\frac{2\operatorname{Area}\bigl({\partial D_{T_1}\cap\partial D_{T_2}}\bigr)}{4G_N},
\end{equation}
where \(I_{\rm bulk}(D:D'):=S_{\rm bulk}(D)+S_{\rm bulk}(D')-S_{\rm bulk}(D\cup D')\ge 0\) is the bulk mutual information between disjoint matter regions. When no shared interface exists, only the mutual-information subtraction appears. Any positive interface area is counted twice in the individual entropies and subtracted when the regions are merged. Since the mutual information and the interface area are nonnegative, \eqref{eq:q-pure-union-claim} together with \eqref{eq:q-pure-mutual-interface} implies the even stronger individual-sum inequality
\begin{equation}\label{eq:q-sum-with-mutual-info}
S_{\rm gen}[T_1]+S_{\rm gen}[T_2]
\ge S_{\rm gen}[T_p]+I_{\rm bulk}(D_{T_1}:D_{T_2})+\frac{2\operatorname{Area}\bigl({\partial D_{T_1}\cap\partial D_{T_2}}\bigr)}{4G_N}.
\end{equation}
\end{remark}

\section{Discussion}
\label{sec:discussion}

Our results establish a quantitative link between geometric decoding obstructions and the entanglement shadow. For a fixed throat with no smaller interior throat (Theorems~\ref{thm:cl-fixed} and~\ref{thm:q-fixed}), an interior bulge exceeding twice the throat area or generalized entropy forces the entire throat interior into the entanglement shadow. For point decoding (Theorems~\ref{thm:cl-point} and~\ref{thm:q-point}), any bulk point whose Python's Lunch barrier exceeds the scale of its outer throat cannot lie on a globally minimizing RT surface or QES unless the geometry realizes the \emph{throats-on-both-sides} configuration. Our core theorems are independent of the Python's Lunch conjecture; only their interpretation as unitary decoding bounds relies on it. Under that dictionary, bulk degrees of freedom that are exponentially hard to decode generically hide inside the entanglement shadow, beyond the reach of boundary entanglement probes.

Inside the entanglement shadow, the exclusion of probe RT surfaces has a direct perturbative consequence. Classically, an infinitesimal metric perturbation compactly supported within the entanglement shadow leaves the areas of all globally minimizing probe RT surfaces unchanged, so boundary subregion entropies exhibit no response. When semiclassical quantum corrections are included, the same insensitivity holds for a perturbative redistribution of matter fields inside the quantum entanglement shadow.\footnote{Concretely, this is realized by an infinitesimal unitary supported strictly within a connected component of the quantum entanglement shadow, accompanied by any metric perturbation confined to that component. Because every globally minimizing entanglement wedge contains either the entire connected component or none of it, such a unitary preserves its bulk entropy. Combined with the unchanged area term, QES stationarity then gives $\delta S(A_1)=0$. Note that smallness of the perturbation alone would not ensure this insensitivity for a general variation of the bulk quantum state.} Perturbatively, boundary entanglement entropies cannot resolve the internal structure of matter deep inside the shadow. This connects to a long-standing question in holography: how is the internal geometry of the entanglement shadow encoded in the dual boundary theory? Concepts such as entwinement~\cite{BalasubramanianEtAl2014} were introduced to address this question, though their explicit boundary realization is currently understood only in specific models with additional symmetries. Our results suggest a complementary operational perspective: information that is exponentially complex to decode via unitary operations may manifest in the bulk as structure inside the entanglement shadow. Of course, what we derive is only a sufficient condition for an entanglement shadow to form; geometries such as $\mathrm{AdS}_3$ conical defects develop an entanglement shadow without requiring a Python's Lunch.

We also showed that the throats-on-both-sides exception pinpoints the only obstruction to an unconditional shadow theorem. If a globally minimizing RT surface or QES passes through a point above the stated barrier threshold, Theorems~\ref{thm:cl-point} and~\ref{thm:q-point} require the existence of two nonempty stopping throats on opposite sides of that probe, satisfying both the individual and joint size constraints while admitting no common admissible cutoff. Importantly, this configuration is a necessary geometric condition for probe penetration rather than a sufficient one: not every two-throat geometry produces a large decoding barrier. In a single-boundary setting, such a throats-on-both-sides configuration resembles multiple black holes enclosed behind a common outer horizon. However, when viewed on a time-reflection-symmetric slice, this spatial bifurcation does not by itself describe an ordinary dynamical black hole merger, since a single enclosing outer horizon is already present at the moment of time symmetry. Understanding the physical origin and dual boundary interpretation of such exceptional configurations remains an interesting open question.

A first limitation of our main theorems is the assumption of a pure boundary state (\(S(A)=0\), \(\gamma_A=\varnothing\)), whereas finite \(S(A)\) arises both for proper subsystems and for the maximally mixed code state \(\rho_{\rm code}\) of \eqref{eq:pl-code-sgen}. At finite \(S(A)\), even the fixed-throat argument requires more than simply excluding smaller interior throats: one must ensure that both cut pieces can monotonically contract all the way to \(\gamma_A\) without halting prematurely. Working in a purification, let \(D_{\gamma_A}\) denote the closed complement of \(\operatorname{EW}(A)\), which lies inside the retained region \(D_T\) and satisfies \(S_{\rm gen}(D_T)=S_{\rm gen}[T] \ge S_{\rm gen}(D_{\gamma_A})=S(A)\). By entanglement wedge nesting, \(\operatorname{EW}(A_1)\subseteq\operatorname{EW}(A)\), so among the two cut pieces in \eqref{eq:q-cut-pieces}, \(D_1\) excludes the region \(D_{\gamma_A}\) whereas \(D_2\) retains it (even if \(\gamma_{A_1}\) touches or shares a segment of the surface \(\gamma_A\), so that \(\gamma_A\) bounds both sides of the shell \(D_T\setminus D_{\gamma_A}\)). QES minimality and strong subadditivity then yield \(S_{\rm gen}(D_1)\le S_{\rm gen}(D_T\setminus D_{\gamma_A})\le S_{\rm gen}[T]+S(A)\) and \(S_{\rm gen}(D_2)\le S_{\rm gen}[T]\). Suppose that for every penetrating probe, \(D_1\) and \(D_2\) admit the monotone contractions of Theorem~\ref{thm:q-fixed} in their respective comparison classes --- shrinking \(D_1\) to \(\varnothing\) and \(D_2\) to \(D_{\gamma_A}\) (so both sides of \(D_T\setminus D_{\gamma_A}\) collapse onto \(\gamma_A\)). Their union then sweeps from \(T\) to \(\gamma_A\) with peak at most \(2S_{\rm gen}[T]+S(A)\), yielding the conditional fixed-throat criterion
\[
S_{\rm gen}\!\bigl[B[T,\gamma_A]\bigr] > 2S_{\rm gen}[T]+S(A)
\quad\Longrightarrow\quad
\operatorname{Int}(D_T\setminus D_{\gamma_A})\subseteq\mathcal S_{\rm ent}^{\rm Q}(A),
\]
which reduces classically to \(\operatorname{Area}(B[T,\gamma_A]) > 2\operatorname{Area}(T)+\operatorname{Area}(\gamma_A)\). Thus, the absence of intermediate stopping surfaces must be imposed separately for regions disjoint from \(D_{\gamma_A}\) (below \(S_{\rm gen}[T]+S(A)\)) and for regions retaining \(D_{\gamma_A}\) (below \(S_{\rm gen}[T]\)). Even when \(\gamma_A=\varnothing\), the external purifier stays with \(D_2\), so a surface that halts the contraction of \(D_1\) need not be a quantum throat with respect to the exterior generalized entropy.

For point decoding, however, raising the threshold to \(\Delta[p]/(4G_N)>S_{\rm gen}[T_p]+S(A)\) is not enough to extend Theorem~\ref{thm:q-point}. Because \(D_1\) is contracted without \(D_{\gamma_A}\), a stable remnant \(D_{T_1}\) of the contraction must still bound an admissible cutoff once \(D_{\gamma_A}\) is adjoined (i.e., \(\partial(D_{T_1}\cup D_{\gamma_A})\) must be an admissible cutoff). Its entropy bound \(S_{\rm gen}[T_p]+S(A)\) alone guarantees neither local minimality under variations carrying \(D_{\gamma_A}\) nor outer-minimality against all admissible enclosures, yet both properties are required for the cutoff-replacement Lemma~\ref{lem:q-cutoff-replace} and the common-enclosure comparison in Theorem~\ref{thm:q-point}.

A second limitation is our restriction to a fixed spatial slice. Our cutting-and-contraction proof relies on partitioning a common Cauchy slice along the probe surface and applying strong subadditivity to the resulting spatial subregions. In a general time-dependent spacetime, the relevant HRT surface or QES and the candidate throats need not lie on a common Cauchy slice, so these slice-wise comparisons do not carry over directly. Consequently, our current slice-wise theorems do not immediately apply to dynamical black hole evaporation. Moreover, even in spacetimes with a time-reflection-symmetric slice, the relevant bulge need not lie on that slice: explicit examples featuring timelike-separated extremal surfaces motivate a covariant maximinimax Python's Lunch prescription~\cite{EngelhardtPeningtonShahbaziMoghaddam2023}. This also raises a broader conceptual question: because the circuit model that originally motivated the Python's Lunch conjecture is formulated via a spatial tensor network on a fixed slice, its operational meaning in a genuinely time-dependent geometry is less clear. Even if our geometric proof extends to the covariant prescription, it remains open whether a covariant Python's Lunch corresponds to an analogous unitary decoding protocol governed by the covariant generalized-entropy barrier.

Testing this relationship in the dual boundary theory will clarify the operational limits of these bounds. The Python's Lunch complexity dictionary models the bulk-to-boundary isometry across a bulge as a generic scrambling circuit, and Grover's search algorithm is merely the fastest general-purpose unitary protocol currently known. In specific holographic theories or structured states, more efficient decoding protocols may exist. From this perspective, the Python's Lunch expression characterizes the worst-case scrambling cost and bounds the true minimal unitary decoding complexity $\mathcal C_{\rm exact}$, provided the encoding and reflection operations can be implemented in the boundary circuit model. In particular, in our throats-on-both-sides exception, it remains open whether reconstructing $p$ truly requires a decoding complexity as large as Definition~\ref{def:classical-point-complexity}. If a simpler boundary protocol achieves $\mathcal C_{\rm exact} < \mathcal C[p]$, our complexity bounds outside the entanglement shadow become even more conservative when expressed in terms of $\mathcal C_{\rm exact}$. While formulating a sharp boundary test will require specifying a concrete decoding task --- gate set, target fidelity, and code subspace --- together with a boundary diagnostic for throat area or generalized entropy, our bulk theorems already establish a clear geometric principle: whenever a geometry avoids the two-sided throat obstruction, high unitary decoding complexity forces bulk information into the entanglement shadow.

\acknowledgments
ChatGPT assisted in the research, and Gemini helped to polish the manuscript. The authors independently verified all derivations and results and take full responsibility for the content of this paper.
This work was supported by the National Natural Science Foundation of China Grants No.~12375041 and~12575046.

\end{document}